\documentclass[esd]{copernicus}

\begin{document}

\title{Using network theory and machine learning \hack{\break} to predict El Ni\~no}

\Author[1,3]{Peter~D.}{Nooteboom}
\Author[1,3]{Qing~Yi}{Feng}
\Author[2]{Crist\'{o}bal}{L\'{o}pez}
\Author[2]{Emilio}{Hern\'{a}ndez-Garc\'{\i}a}
\Author[1,3]{Henk~A.}{Dijkstra}

\affil[1]{Institute for Marine and Atmospheric Research Utrecht (IMAU), Department of Physics, \hack{\break} Utrecht University, Utrecht, the Netherlands}
\affil[2]{Instituto de Física Interdisciplinar y Sistemas Complejos (IFISC, CSIC-UIB), University of the Balearic Islands, Balearic Islands, Spain}
\affil[3]{Centre for Complex Systems Studies, Utrecht University, Utrecht, the Netherlands}

\runningtitle{Using machine learning to predict El Niño}

\runningauthor{P.~D.~Nooteboom et al.}

\correspondence{Peter~D.~Nooteboom (p.d.nooteboom@uu.nl)}

\received{7 March 2018}
\pubdiscuss{13 March 2018}
\revised{22 June 2018}
\accepted{26 June 2018}
\published{}

\firstpage{1}

\texlicencestatement{This work is distributed under \hack{\newline} the Creative Commons Attribution 4.0 License.}
\maketitle

\begin{abstract}
The skill of current predictions of the warm phase of the El Niño Southern
Oscillation (ENSO) reduces significantly beyond a lag time of 6~months. In
this paper, we aim to increase this prediction skill at lag times of up to
1~year. The new method combines a classical autoregressive integrated moving
average technique with a modern machine learning approach (through an
artificial neural network). The attributes in such a neural network are
derived from knowledge of physical processes and topological properties of
climate networks, and they are tested using a Zebiak–Cane-type model and
observations. For predictions up to 6~months ahead, the results of the hybrid
model give a slightly better skill than the CFSv2 ensemble prediction by the
National Centers for Environmental Prediction (NCEP). Interestingly, results
for a 12-month lead time prediction have a similar skill as the shorter lead
time predictions.
\end{abstract}

\introduction

Approximately every 4~years, the sea surface temperature (SST) is higher
than average in the eastern equatorial Pacific
\citep{Philander1990ElOscillation}. This phenomenon is called an El Niño and
is caused by a large-scale ocean--atmosphere interaction between the
equatorial Pacific and the global atmosphere
\citep{Bjerknes1969AtmosphericPacific}, referred to as the El Niño Southern
Oscillation (ENSO). It is the dominant mode of climate variability at
interannual timescales and has teleconnections worldwide. As El~Ni\~no
events cause enormous damage worldwide, skillful predictions, preferable for
lead times up to 1~year, are highly desired.

So far, both statistical and dynamical models are used to predict ENSO
\citep{Chen2004PredictabilityYears.,Yeh2009ElClimate,Fedorov2003HowNino}.
However, El~Niño events are not predicted well enough up to 6~months ahead
due to the existence of the so-called predictability barrier
\citep{Goddard2001CurrentPredictions}. Some theories indicate that this is due to
the chaotic, yet deterministic, behaviour of the coupled atmosphere--ocean
system \citep{Jin1994ElChaos,Tziperman1994ElOscillator}. Others point out the
importance of atmospheric noise, acting as a high-frequency forcing
sustaining a damped oscillation \citep{Moore1999StochasticOscillation}.

Recently, attempts have been made to improve the ENSO prediction skill beyond
this spring predictability boundary, for example by using machine learning
(ML; \citealp{Wu2006NeuralTemperatures}) methods, also combined with network
techniques \citep{Feng2016ClimateLearn:Measures}. ML has shown to be a
promising tool in other branches of physics, outperforming conventional
methods \citep{Hush2017MachinePhysics}. As the amount of data in the climate
sciences is increasing, ML methods such as artificial neural networks (ANNs),
are becoming more interesting to apply to prediction studies.

Briefly, ANN is a system of linked neurons that describes, after
optimization, a function from one or more input variables (or attributes) to
the output variable(s). Generally, one has to choose how large and
complicated the ANN structure is. The more complicated an ANN, the more it
will filter the important information from the attributes itself, but it will
require more input data and is computationally intensive. Therefore, simpler
ANN structures are used in this article. However, techniques will have to be
applied in order to reduce the amount of input variables and select the
important ones, to make the problem appropriate for the simpler ANN. This
reduction and selection problem can be tackled in many ways, which are
crucial for the prediction. The main issue in these methods, however, is what
attributes to use for ENSO prediction.

Complex networks turn out to be an efficient way to represent spatiotemporal
information in climate systems
\citep{Tsonis2006WhatClimate,Steinhaeuser2012MultivariateNetworks,Fountalis2015ENSOCentury}
and can be used as an attribute reduction technique. These climate networks
are in general constructed by linking spatiotemporal locations that are
significantly correlated with each other according to some measure. It has
been demonstrated that relationships exist between topological properties of
climate networks and nontrivial properties of the underlying dynamical system
\citep{Deza2014DistinguishingNetworks,Stolbova2014TopologyLanka}, also
specifically for ENSO
\citep{Gozolchiani2011EmergenceNetwork,Gozolchiani2008PatternEvents,Wang2015OceanicNetworks}.
Climate networks already appear to be a useful tool for more qualitative
ENSO prediction, by considering a warning of the onset of El~Niño when a
certain network property exceeds some critical value
\citep{Ludescher2014VeryNino.,Meng2017PercolationConditions,Rodriguez-Mendez2016Percolation-basedSystems}.

In this paper, a hybrid model is introduced for ENSO prediction. The model
combines the classical linear statistical method of autoregressive integrated
moving average (ARIMA) and an ANN method. ANN is applied to predict the
residual, due to the nonlinear processes, that is left after the ARIMA
forecast \citep{Wu2006NeuralTemperatures}. To motivate our choice for
attributes in the ANN, we use an intermediate-complexity model which can
adequately simulate ENSO behaviour, the Zebiak--Cane (ZC) model
\citep{Zebiak1987AOscillation}. The attributes which are used in the
prediction model are related to physical processes which are relevant for
ENSO prediction. Moreover, network variables are considered as attributes
such that they relate to a physical mechanism, but additionally contain
spatial information.

Section~2 briefly describes the ZC model, the methods considering both the
climate networks and ML, and the data from observations. In Sect.~3, the
network methods are first applied to the ZC model. Second, the attributes
selected for observations are presented. These attributes, among which there
is a network variable, are applied in the hybrid prediction model in Sect.~4,
which discusses the skill of this model to predict El Ni\~no. The paper
concludes with a summary and discussion in Sect.~5.

\section{Observational data, models and  methods\label{secmod}}

\subsection{Data from observations}

As observational data, we use the sea surface height (SSH) from the weekly
ORAP5.0 (Ocean ReAnalysis Pilot 5.0) reanalysed dataset of ECMWF from 1979 to
2014 between 140 to $280{\degree}$\,E and $20{\degree}$\,S to
$20{\degree}$\,N.

For recent predictions, the SSALTO/DUACS altimeter products are used for the
same spatial domain, since the SSH is available from 1993 up to the present
in this dataset. The SSALTO/DUACS altimeter products were produced and
distributed by the Copernicus Marine and Environment Monitoring Service
(\url{http://marine.copernicus.eu/}, June 2017).

In addition, the HadISST dataset of the Hadley Centre has been used for the SST and the
NCEP/NCAR Reanalysis dataset for the wind stress from 1980 to the present
\citep{Rayner2003GlobalCentury}.

To quantify ENSO, the NINO3.4 index is used, i.e. the 3-month running
mean of the average SST anomaly in the extended reconstructed SST dataset
between 170 to $120{\degree}$\,W and $5{\degree}$\,S and
$5{\degree}$\,N \citep{Huang2015ExtendedIntercomparisons}.

The warm water volume (WWV), being the integrated volume above the
$20\,{\degree}$C isotherm between $5{\degree}$\,N--$5{\degree}$\,S and
120--280{\degree}\,E, is determined from the temperature analyses of the
Bureau National Operations Centre
(\url{https://www.pmel.noaa.gov/elnino/upper-ocean-heat-content-and-enso},
\citealp{NOAA}).

\subsection{The Zebiak--Cane model}

\begin{figure*}[t]
\includegraphics[width=12cm]{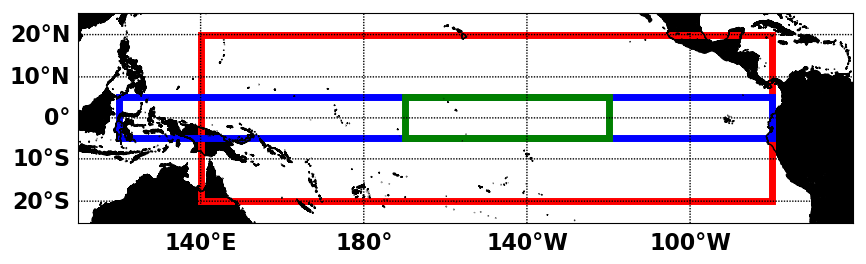}
\caption{Pacific area (red rectangle) from 140--280{\degree}\,E to
$20${\degree}\,S--20{\degree}\,N, the NINO3.4 area (green rectangle) from
170--120{\degree}\,W to $5${\degree}\,S--5{\degree}\,N and the WWV area
(blue rectangle) from 120--280{\degree}\,E to
5{\degree}\,S--5{\degree}\,N.} \label{fig:pac}
\end{figure*}

The ZC model \cite[]{Zebiak1987AOscillation} represents the coupled
ocean--atmosphere system on an equatorial $\beta$-plane in the equatorial
Pacific (see Fig.~\ref{fig:pac}). This model is used here to infer which
processes are important for ENSO prediction and to find the attributes which
represent those processes. Also, a network analyses is applied to the ZC
model in order to find network variables which could improve prediction,
before these network variables are calculated in observations. We use the
numerically implicit version of this model
\citep{vanderVaart2000TheModel,VonDerHeydt2011ColdPliocene} as in
\cite{Feng2015AVariability}.

In the ZC model, a shallow-water ocean component is coupled to a steady
shallow-water Gill atmosphere model \cite[]{Gill1980SomeCirculation}. The
atmosphere is driven by heat fluxes from the ocean, depending linearly on the
anomaly of the SST $T$ with respect to a radiative
equilibrium temperature $T_0$. The zonal wind stress $\tau^x$ is the sum of a
coupled and an external part. The external part is independent of the
coupling between the atmosphere and ocean and represents a weak easterly wind
stress due to the Hadley circulation. The coupled part of the zonal wind
stress is proportional to the zonal wind from the atmospheric model; the
meridional component of the wind stress is neglected in this model.

As shown in \cite{vanderVaart2000TheModel}, the parameter measuring the
magnitude of the ocean--atmosphere coupled processes is the coupling strength
$\mu$. Without any included noise, a temperature anomaly damps out to a
constant value and a stationary state if $\mu<\mu_\mathrm{c}$, where
$\mu_\mathrm{c}$ indicates a critical value. However, if the coupling
strength exceeds the critical value $\mu_\mathrm{c}$, a supercritical Hopf
bifurcation occurs. A perturbation then does not decay, but an oscillation is
sustained with a period of approximately 4~years.

Three positive feedbacks related to the thermocline depth, upwelling and
zonal advection can cause the amplification of SST anomalies
\citep{Dijkstra2006TheMechanisms}, while the oscillatory behaviour associated
with ENSO is caused by negative delayed feedbacks. The ``classical delayed
oscillator'' paradigm assumes this negative feedback is caused by waves
through geostrophic adjustment, controlling the thermocline depth. A
complementary, different view is the ``recharge/discharge oscillator''
\citep{Jin1997AnModel}, also regarding oceanic waves excited through oceanic
adjustment. The waves excited to preserve the Sverdrup balance are
responsible for a transport of warm surface water to higher latitudes,
discharging the warm water in the tropical Pacific. The thermocline depth is
raised, resulting in more cooling of SST. The WWV is the
variable generally used to capture how much the tropical Pacific is
``charged''.

Apart from the coupled ocean--atmosphere processes, ENSO is also affected by
fast processes in the atmosphere, which are considered as noise in the ZC
model. An important example of atmospheric noise are the so-called westerly
wind bursts (WWB). These are related to the Madden--Julian oscillation
\citep{Madden1994ObservationsReview}. The WWB is a strong westerly anomaly in
the zonal wind field, occurring every 40 to 50 days and lasting
approximately 1~week. The effect of the noise on the model behaviour depends
on whether the model is in the super- or subcritical regime (i.e. whether
$\mu$ is above or below $\mu_\mathrm{c}$). If $\mu<\mu_\mathrm{c}$, the noise
excites the ENSO mode, causing irregular oscillations. In the supercritical
regime, a cycle of approximately 4~years is present, and noise causes a
larger amplitude of ENSO variability.

The atmospheric noise in the model is represented by obtaining a residual of
the wind stress from observations as in \cite{Feng2016ClimateVariability}.
Since weekly data are considered, every discrete time step in the model is
1~week.

\subsection{Network variables}

Here we explain the methods to calculate a property of a climate network
which is tested in the ZC model and observations and will be used in the
hybrid model. From the network analysis we found several climate network
quantities with interesting properties for prediction, but which are not used
in the hybrid model of the next section. The methods to calculate these
properties can be found in Appendix~\ref{sec:a1}.

An undirected and unweighted network is constructed making use of the Pearson
correlation of climate variables related to ENSO (e.g. SST, thermocline depth
or zonal wind stress). Network nodes are model or observation grid positions
$i$ and the links are stored in a symmetric adjacency matrix $A$, where
$A_{ij}=1$ if node $i$ is connected to node $j$ and $A_{ij}=0$ otherwise.
$A_{ij}$ is defined by
\begin{equation}
A_{ij}=\Theta\left(\left|R_{ij}\right|-\epsilon\right)-\delta_{ij} .
\end{equation}
Here $R_{ij}$ is the Pearson correlation between node $i$ and $j$,  $\epsilon$ is
the threshold value and $\Theta$ denotes the Heaviside function. Hence, if the Pearson
correlation exceeds the threshold $\epsilon$, the two nodes  will be linked.  The $\delta_{ij}$ is the
Kronecker delta function, implemented to prevent connection of nodes with themselves.

Percolation theory is then considered, describing the connectivity of
different clusters in a network. It has been found that the connectivity of
some climate networks increases just before an El~Niño and decreases
afterwards \citep{Rodriguez-Mendez2016Percolation-basedSystems}, as local
correlations between points increase and decrease. At such a percolation-like
transition, the addition of only a few links can cause a considerable part of
the network to become connected. Before the percolation transition, clusters
of small sizes will form. Therefore the variable $c_s$ will warn for the
transition:
\begin{equation}
c_s = \frac{sn_s}{N}.
\end{equation}
Here $n_s$ is the amount of clusters of size $s$ and $N$ the size (i.e. the total amount of nodes) of the network.
Thus $c_s$ is the fraction of nodes that are
part of a cluster of (generally small) size $s$.

\subsection{Hybrid prediction model \label{secml}}

A hybrid model \citep{Valenzuela2008HybridizationPrediction} will be applied to predict
ENSO, in which the observation $Z_t$ at time $t$ is represented by
\begin{equation}
Z_t= Y_t + N_t .\label{eq:add}
\end{equation}
Here $Y_t$ is modelled by a  linear process and $N_t$ by a ML-type technique.  Let $\tilde{Y}_t$ be
the prediction of the part $Y_t$ using ARIMA, then  $Z_t-\tilde{Y}_t$ is the residual with respect to
the observed value.  This  residual will be predicted by the feed-forward ANN:
\begin{equation}
\tilde{N}_t = f\left(x_{1}(t),\cdots,x_{N}(t)\right) .
\end{equation}
Here $f$ is a nonlinear function of the $N$ attributes
$x_{1}(t),\cdots,x_{N}(t)$ and $\tilde{N}_t$ the prediction of residual
$Z_t-\tilde{Y}_t$ at time $t$. Notice the nonlinear function $f$ does not
depend on history, whereas the ARIMA part $\tilde{Y}$ does. The final
prediction $\tilde{Z}_t$ of the hybrid model is
\begin{equation}
\tilde{Z}_t=\tilde{Y}_t + \tilde{N}_t .
\end{equation}
Previous work showed that the results of a hybrid model are in general more
stable and reduce the risk of a bad prediction, compared to a single
prediction method \citep{Hibon2005ToCombinations}. ``More stable'' means that
a hybrid model has a lower variability in prediction skill for different
arbitrary time series. Besides, ARIMA is a simple method to include
information about the history in the prediction model, which is not in the
feed-forward ANN.

This scheme describes a ``supervised'' model, implying that the predictant is
``known''. This known quantity is the NINO3.4 index. The standard procedure
for supervised learning is to optimize the ML method on a ``training set'' to
define an optimal model, which predicts ENSO with a certain time ahead. This
function will then be tested on a test set. Here a training set of $80\,\%$
and a test set of $20\,\%$ of the total time series is used. The dataset can
be represented by a $T\times N$ matrix, where $T$ represents the length of
the time series and each time $t=1,_{\cdots},T$ has a set of $N$ attributes
$x_1(t),_{\cdots},x_N(t)$. Note that, since we are predicting time series,
for any training set $[t_i^\mathrm{train},t_f^\mathrm{train}]$ and test set
$[t_i^\mathrm{test},t_f^\mathrm{test}]$,
$t_i^\mathrm{test}>t_f^\mathrm{train}$ is convenient (where
$t_i^\mathrm{train},t_f^\mathrm{train},t_i^\mathrm{test},t_f^\mathrm{test}\in[1,T]$).
In the following, we describe more in detail the different parts of this
hybrid prediction method.

First, the training set is used to optimize an ARIMA($p,d,q$) process for
the NINO3.4 time series. The standard method maximizing the log likelihood
function is used to fit $\alpha_1,\cdots,\alpha_{p}, \beta_1,\cdots,\beta_q$,
such that $\sum_t\varepsilon_t^2$ is minimized for time series $Z_t$ with $t$
in months:
\begin{align}
& \left( 1-\alpha_1 B-_{\cdots}-\alpha_pB^p\right)\left( 1-B\right)^d Z_t
\nonumber\\
& ~~~= \left(1 + \beta_1B+_{\cdots}+\beta_qB^q\right)\varepsilon_t, \label{eq:arima}
\end{align}
where $\varepsilon_t$ is the residual, differencing order $d$ determines the
amount of differencing terms, $p$ is the amount of autoregressive terms and
$q$ is the amount of moving average terms on the right-hand side; $B$ ($B
Z_t=Z_{t-1}$) is the lag operator. Finding the most optimal ARIMA order
$(p,d,q)$ is not trivial
\citep{Zhang2003TimeModel,Aladag2009ForecastingMethodology}. General methods
include the Akaike's information criterion \citep{Akaike1974AIdentification}
or minimum description length \citep{Rissanen1978ModellingDescription}.
However, these methods are often not satisfactory and additional methods have
been proposed to determine the order \citep{Al-Smadi2005ARMAPerspective}. In
this article we mainly present results obtained with orders $p=12$, $d=1$ and
$q=0$ or $q=1$, which gave good prediction skill and it can be argued that in
such a chaotic system, information from too long ago is not important
anymore.

The eventual ARIMA equation results in a prediction
$\hat{Y}_t(Z_{t-1},_{\cdots},Z_{t-p},\varepsilon_{t-1},_{\cdots},\varepsilon_{t-q})$
of $\tau=1$ months ahead. Here $\varepsilon_{t-1}=Z_{t-1}-\hat{Y}_{t-1}$. Let
$\tilde{Y}_t$ be the ARIMA prediction of $\tau>0$ months ahead, by
calculating $\hat{Y}_t$ for $\tau$ times in the future and replacing any
observation $Z_{t}$ with the consecutive calculated $\hat{Y}_t$, where $t$ is
in the future and $Z_{t}$ therefore unknown. Similarly, if $q=1$ and $\tau>1$
months, the residual is calculated by
$\epsilon_{t-1}=\tilde{Y}_{t-1}-\hat{Y}_{t-1}$, since the observed value
$Z_{t-1}$ is in the future. Hence the ARIMA prediction $\tilde{Y}_t$ will be
a time extrapolation with the optimized ARIMA model.

After $\tilde{Y}_t$ is predicted by the ARIMA model, the ANN will be used for
the prediction $\tilde{N}_t$, making use of more variables than the NINO3.4
index alone. Deciding which of the variables to use is not a straightforward
problem, yet crucial for the eventual prediction. Generally in an ANN, a pair
of two variables can be compatible in the prediction, but perform poor when
applied alone. Other pairs can be redundant and cover important information
when used alone, but solely noise is included when used together
\citep{Guyon2003AnSelection}. Adding a variable to the attribute set and
seeing if it improves prediction can only conclude whether it improves prediction
with respect to the old attribute set, not whether the variable is predictive
in itself. To determine the attribute set, we consider which variables
represent a certain physical mechanism that is important for the ENSO
prediction. This helps to find attributes which are not related to each
other, but include important information on their own. Besides, it is tested
whether the prediction skill is reduced if a variable is dropped out of the
attribute set.

Moreover, at every lead time an optimal attribute must be selected. Hence the
final prediction model is tuned for a specific lead time and will not be a
step by step prediction forward in time. Apart from considering the physical
mechanisms the variables represent, two methods will help to decide which
variables can improve the prediction. First, correlation between the
predictor and predictant is a commonly used measure for attribute selection
\citep{Hall1999Correlation-basedLearning}. Therefore the Pearson
cross-correlation is calculated for the attributes at lag $\tau$ to show the
predictability of a time series:
\begin{equation}
\hack{\hbox\bgroup\fontsize{9.5}{9.5}\selectfont$\displaystyle}
R_{\tau}(p,q)=\max_{\tau}\left(\frac{\sum^{n}_{k=1}p(t_k)q(t_k-\tau)}
{\sqrt{\left(\sum^{n}_{k=1}p^2(t_k)\right)\left(\sum^{n}_{k=1}q^2(t_k-\tau)\right)}}\right)
.\label{eq:crosscor}
\hack{$\egroup}
\end{equation}
Here $p$ is the predictor, $q$ is the predictant and lag $\tau\leq64$ weeks
such that no information too far in the past is considered.

However, the effect of a variable on ENSO at a short lead time increases the
cross-correlation at a longer lead time, due to the effect of autocorrelation
\citep{Runge2014DetectingSystems}. To solve this autocorrelation problem, a
Wiener--Granger causality $F$~test \citep{Sun2014UsingSeries} is performed
between all predictors $x_1,\cdots,x_N$ and the predictant at lags $\tau$.
Note Granger causality is not the same as a ``true'' causality. If the test
results in a low $p$~value, the null hypothesis that $x_i$ does not
cause in the Granger sense the predictant,
due to Granger causality, is rejected at a low significance level (i.e. $x_i$ is
more likely to cause the predictant due to Granger causality). Notice that both the
cross-correlation and Wiener--Granger method give us merely an idea of which
variables can be used for the prediction at different lead times. Both
methods are linear, while the attributes will be used in a nonlinear method.

Finally, the $T\times N$ dataset with selected attributes is used to predict
the residual between the ARIMA forecast and the observations in an ANN.
Besides using the NINO3.4 sequence itself, the additional attributes can be
applied to add important information and improve the prediction.

In this paper, only a feed-forward ANN is applied, having a structure without
loops. The input variables are linearly combined and projected to the first
layer neurons according to \citep{Bishop2006PatternLearning}:
\begin{equation}
z_j=h\left(\sum^{D}_{i=1}w^{(1)}_{ji}x_i+w^{(1)}_{j0}\right) .
\end{equation}
Here $z_j$ is the value of the $j$th neuron of the layer; $w^{(1)}_{ji}$ is
the weight between input $x_i$ from neuron $i$ to neuron $j$, where the
(1)~denotes the first layer. $w^{(1)}_{j0}$ is referred to as the bias. $h$
is the sigmoid activation function, essential for incorporating the
nonlinearity in the prediction model.

These $z_j$ can again be used as input for a second layer, which can be used
for a third layer, and so on. Eventually this leads to some output which can be
compared with the time series that must be predicted. Using a backward-propagating
technique, the squared error $\sum_t(y_t-\hat{y}_t)^2$ between
the residual we are predicting $y_t$ and the output of the ANN $\hat{y}_t$
will be minimized over the weights for the training set. The optimized
function can then be tested on the test set. Initially, some random
distribution of weights is used. The ANN part of the prediction will be
performed with the toolbox ClimateLearn
\citep{Feng2016ClimateLearn:Measures}.

To summarize the tuning of the hybrid model: the ARIMA order and the hyperparameters
controlling the ANN structure are tuned on the data, i.e. such that the prediction
result is optimal. However, we will consider whether some set of different parameter
values converges to similar predictions, which can show whether the hyperparameter
tuning was a one lucky shot or not. The choice of the attributes is based on the ZC model
giving a more physical basis for the information needed for a good prediction. To select
them at a specific lag their cross-correlation and Wiener--Granger causality with the
ENSO index and performance are also considered, which could lead to the replacement of an
attribute with another attribute which is physically related.

\section{Analysis of network properties and selection of ML attributes}

In this section, topological properties of climate networks are analysed within the ZC model and observations,
which lead  to specific choices of attributes in the hybrid prediction model.

\subsection{Network variables from the ZC model}

Weekly spatiotemporal data
on a $31\times30$ grid in the Pacific region are
obtained for 45~years from the ZC model, to construct the climate
networks. The first 5~years are not considered, to discard the effect of
the initial conditions. A sliding-window approach is used to calculate the
network variables. This implies that a different network is calculated at
each time, which is sliding 4~weeks ahead every time step. For the ZC
model, either the thermocline network (from $h$), SST network (from $T$),
wind-stress network (from $\tau^x$) or a combination of these are considered
for network construction. Only the network variable which showed the same
behaviour in the observations and in the ZC model is presented here. Other
network variables with interesting properties can be found in
Appendix~\ref{sec:a2}.

\begin{figure*}[t]
      \includegraphics[width=12cm]{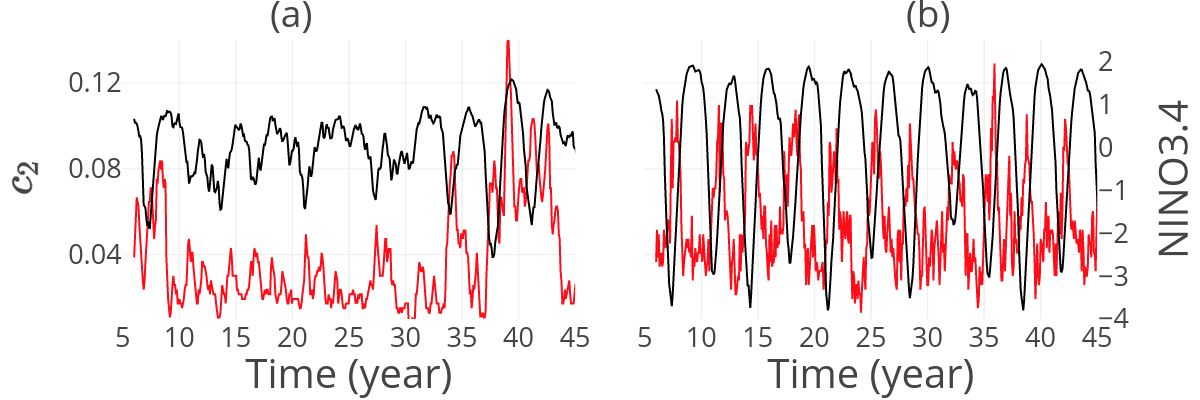}
      \caption{The network variable $c_2$ of the thermocline network with a
      sliding window of 1~year in red and NINO3.4 in black in the ZC model.
      \textbf{(a)}~The subcritical ($\mu=2.7$) case with threshold $\epsilon=0.99999$ and \textbf{(b)}~the
            supercritical ($\mu=3.25$) case with $\epsilon=0.999$.}
      \label{fig:c2}
\end{figure*}
The network variable of interest is $c_2$ (the proportion of nodes belonging
to clusters of size two) of the thermocline network, because it indicates the
approach to a percolation transition of the network during an El~Ni\~no event
(Fig.~\ref{fig:c2}). A window of 1~year is used. $c_2$ increases
approximately 1 to 2~years before an El~Niño event. This is mainly clear
in the supercritical case. In the subcritical case, a clear warning of an
event occurs when the oscillation of ENSO is more clear and the El~Niños are
stronger. Because $c_2$ is a warning signal of an El~Ni\~no event in the ZC
model, we will look in the next section at how it behaves when it is calculated
from observations.

\subsection{Selecting attributes from observations\label{secobs}}

The ZC model results have given an indication of the network variables that
could be used as attributes in the hybrid model to predict El~Niño. Although
the network variables show interesting behaviour in the ZC model for
prediction, this is not always the case in observations. This section
describes which variables, including a network variable, are implemented in
the hybrid model and the selection of these attributes at different lead
times. Notice that only anomalies of the time series in observations are
considered.

\begin{figure}[t]
\includegraphics[width=8.5cm]{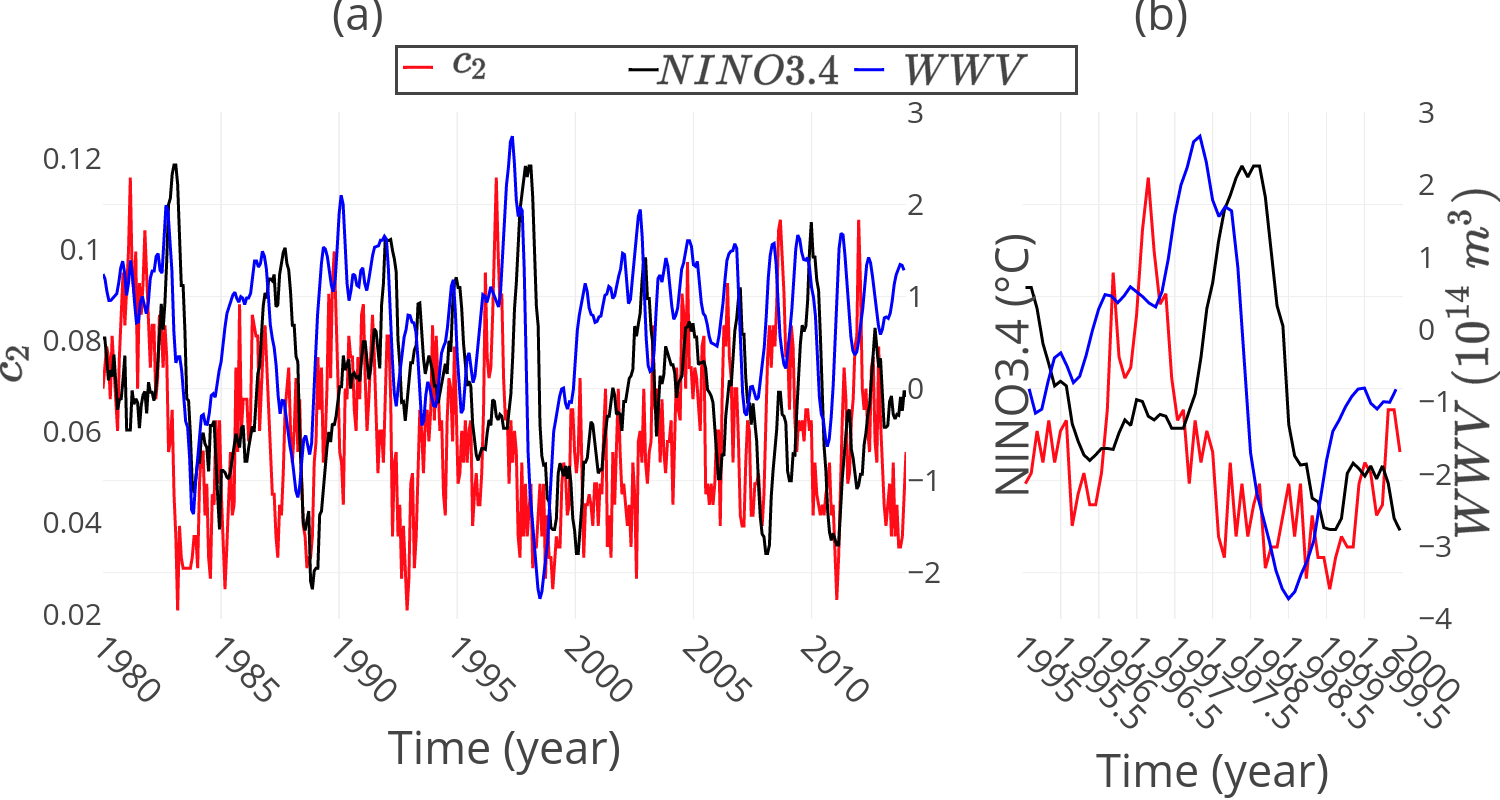}
\caption{The WWV, $c_2$ and the NINO3.4 index from observations for
\textbf{(a)}~the whole considered time series and \textbf{(b)}~only during
the 1997 El~Niño. A warning of the El~Niño event is visible for the WWV and
$c_2$. $c_2$ gives a warning almost a year before the 1997 El~Niño, while
the WWV warns almost 7~months ahead.} \label{fig:c2_obs}
\end{figure}

\begin{figure}[t]
\includegraphics[width=8.5cm]{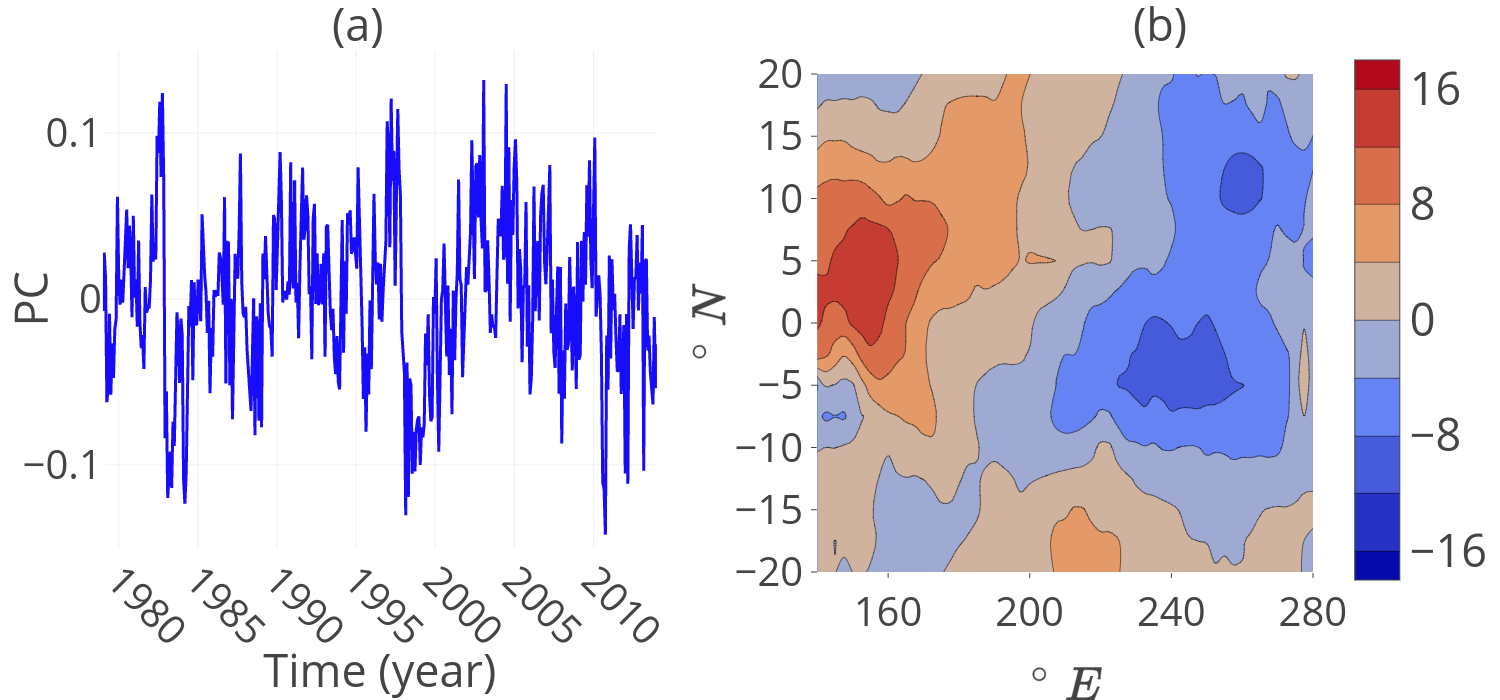}
\caption{\textbf{(a)}~The second principal component of the residual of the
wind stress (PC$_2$) and \textbf{(b)}~its EOF, associated with the WWBs.}
\label{fig:PC}
\end{figure}

First, from the recharge/discharge oscillator point of view, the WWV shows
great potential for the prediction of ENSO
\citep{Bosc2008ObservedOcean,Bunge2014OnENSO}. Therefore it is used in the
attribute set. The second attribute is a network variable related to WWV. The
correlations of the SSH time series on a grid of $27$ latitude points and
$30$ longitude points in the Pacific area are used to reconstruct a network
with a threshold $\epsilon=0.9$ and a sliding window of 1~year. The SSH is used instead of thermocline depth, because more data
is available and it is by approximation proportional to the thermocline depth
\citep{Rebert1985RelationsOcean}. During an El~Niño event, the link density
of this network increases in the warm pool and the cold tongue specifically,
causing a percolation-like transition. As discussed in the previous section,
an early warning could be obtained with $c_2$. This variable allows us to
extend the lead time of the WWV (Fig.~\ref{fig:c2_obs}). Third, atmospheric
noise from the WWBs are a limitation for the prediction of ENSO
\citep{Moore1999StochasticOscillation,Latif1988TheBursts}. To obtain a
variable related to the WWBs, the linear effect of the SST is subtracted from
the zonal component of the wind stress. The second principal component
(PC$_2$), explaining $8\,\%$ of the variance, is associated with these WWBs.
In Fig.~\ref{fig:PC}, the principal component and its empirical orthogonal function (EOF) are presented. The
peaks in the principal component are visible before the great El~Niño events
of $1982$ and $1997$. Thereby, the EOF has the typical WWB structure, being
positive west from the dateline and negative east. Finally, the attribute set
does not yet contain any information about the seasonal cycle (SC) yet. The
phase locking of an El~Niño event to boreal winter is very typical to ENSO.
Therefore a sinusoid with a period of 1~year is used as attribute, to see
if it can improve the prediction skill.

\begin{figure}[t]
\includegraphics[width=8.5cm]{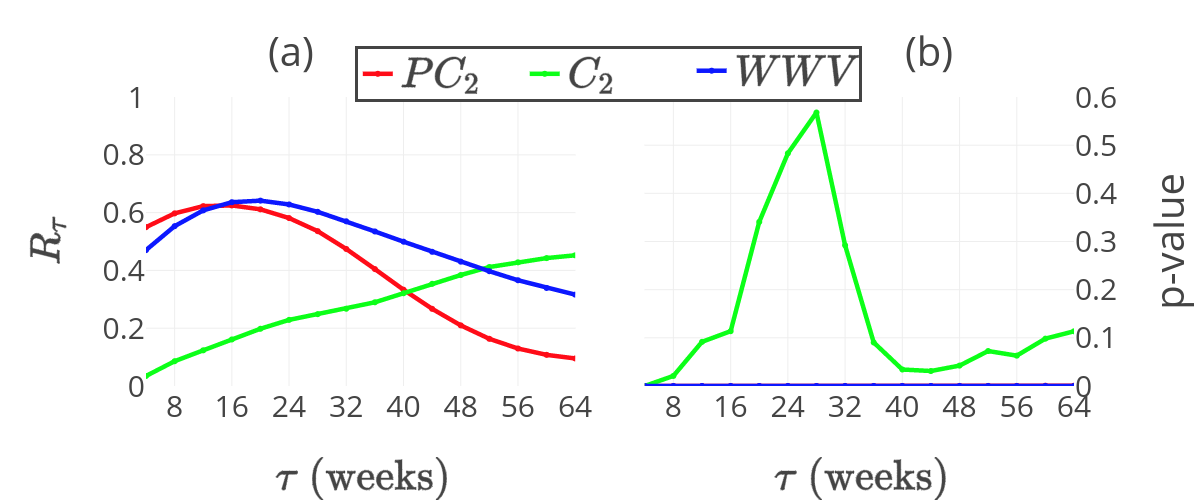}
\caption{\textbf{(a)}~The cross-correlation of the PC$_2$, WWV and $c_2$ with
respect to NINO3.4 for different lags $\tau$. \textbf{(b)}~The $p$~value of
the Wiener--Granger hypothesis test for the same lags. A low $p$~value implies
the variable is likely to cause the NINO3.4 index at the specific
lag due to Granger causality. The $p$~values of the PC$_2$ and WWV are almost zero for all lags.}
\label{fig:CC_WG}
\end{figure}

To determine at which lead time the different attributes should be applied,
the cross-correlation and the $p$~value of the Granger test between the
attributes and NINO3.4 are considered (Fig.~\ref{fig:CC_WG}). The
cross-correlations of PC$_2$ and the WWV show peaks at respectively 12 and 20
weeks, indicating their optimal lead times, since the $p$~values of the
Granger tests are low at every lag and autocorrelation does not play an
important role. For $c_2$, however, the cross-correlation increases up to the
maximum considered lag, but the $p$~value of the Granger test has a local
minimum close to a lag of 44~weeks. According to these methods, $c_2$ is
especially predictive at the longer lead times close to 44~weeks.

To summarize, we are interested in the variables that represent specific
physical characteristics related to the prediction of ENSO, to select the
attributes. Both $c_2$ and the WWV are related to the recharge/discharge
mechanism. PC$_2$ is related to the atmospheric noise from WWBs. The SC is related to the phase locking of El~Niño events to boreal
winter. The hybrid model allows us to implement different variables in the
attribute set at different lead times. Therefore, the cross-correlations and
Wiener--Granger causality were used to determine which attribute is more
optimal at various lead times. This showed that it is better to use $c_2$
instead of WWV at lead times of more than 40~weeks. The other network
variables which were interesting for the ZC model output (see the Appendix)
are performing worse when applied to observations and hence are not used as
attributes in the hybrid model.

\section{Prediction results}

\begin{figure}
      \includegraphics[width=8.5cm]{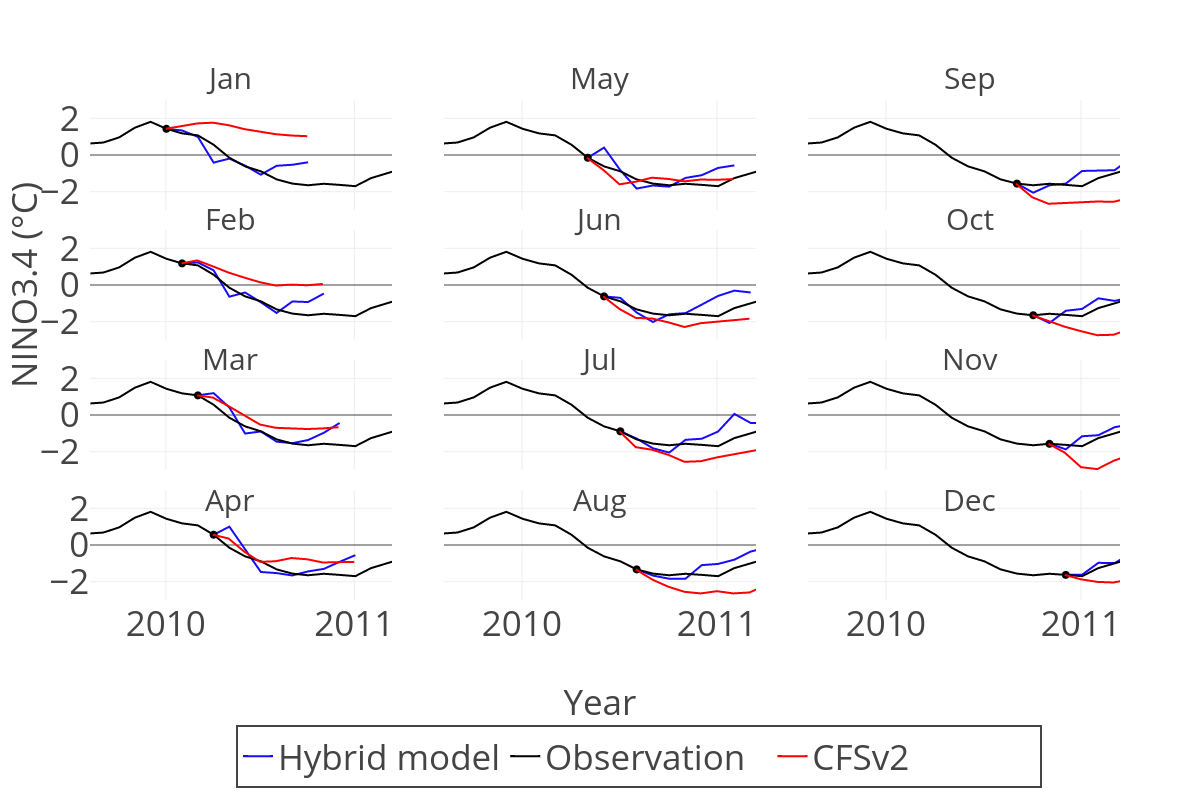}
      \caption{The 9-month ahead prediction starting from every month in the year 2010.
      Blue is the hybrid model prediction with ARIMA(12,1,1),
      $2\times 1\times 1$ ANN structure and attributes are the 3-month running mean of WWV,
      PC$_2$ and SC. The black line is the observed index. Red is the mean of the
      CFSv2 ensemble prediction. \label{fig:2010}}
\end{figure}

This section presents the predictions of the hybrid model, as compared with
observations and with alternative predictions from the CFSv2 model ensemble
of NCEP. The skill with ANN structures up to three hidden layers is
investigated. First, a comparison between both predictions is made for the
year 2010 (Fig.~\ref{fig:2010}). Moreover, several lead time predictions are
shown and compared to the available CFSv2 lead time predictions. Next it is
shown that these prediction models converge to similar results for different
hyperparameters and when using different training and test sets in a
cross-validation method. Finally, a recent forecast is made and it is shown
how the hybrid model predicts the development of ENSO the coming year.

\hack{\newpage}
From now on, the normalized root mean squared error (NRMSE) is used to
indicate the skill of prediction within the test set:
\begin{eqnarray}
\nonumber
\mathrm{NRMSE}(y^A,y^B)=\frac{1}{\max\left(y^A,y^B\right)-\min\left(y^A,y^B\right)}
 \\ \times \sqrt{\frac{\sum_{t_1^\mathrm{test}\leq t_k\leq
t_n^\mathrm{test}}\left(y^A_k-y^B_k\right)^2}{n}} .\label{eq:NRMSE}
\end{eqnarray}
Here $y^A_k$ and $y^B_k$ are respectively the NINO3.4 index and its prediction at
time $t_k$ in the test set. $n$ is the number of points in the test set. A
low NRMSE indicates the prediction skill is better. For all presented
hindcasts, the ARIMA prediction had a significant residual, which implies
that the addition of the ANN part improved prediction.

The year 2010 is a recent example of an under-performing CFSv2 ensemble.
Especially in January, all members of the ensemble overestimate the NINO3.4
index, resulting in an overestimation of the ensemble mean (see
Fig.~\ref{fig:2010}). The hybrid model is used to predict the same period,
with ARIMA(12,1,1) and a $2\times 1\times 1$ ANN structure with the
3-month running mean of the WWV, PC$_2$, the SC and NINO3.4
itself as attributes. In this case the hybrid model performs better than the
CFSv2 ensemble. A $2\times 1\times 1$ structure means a feed-forward
structure with three layers of respectively two, one and one neuron. This ANN
structure is found to be the best performing structure in terms of NRMSE at a
3-month lead time prediction. It will probably not be the most optimal
ANN structure at other lead times.

\begin{figure*}[t]
      \includegraphics[width=11cm]{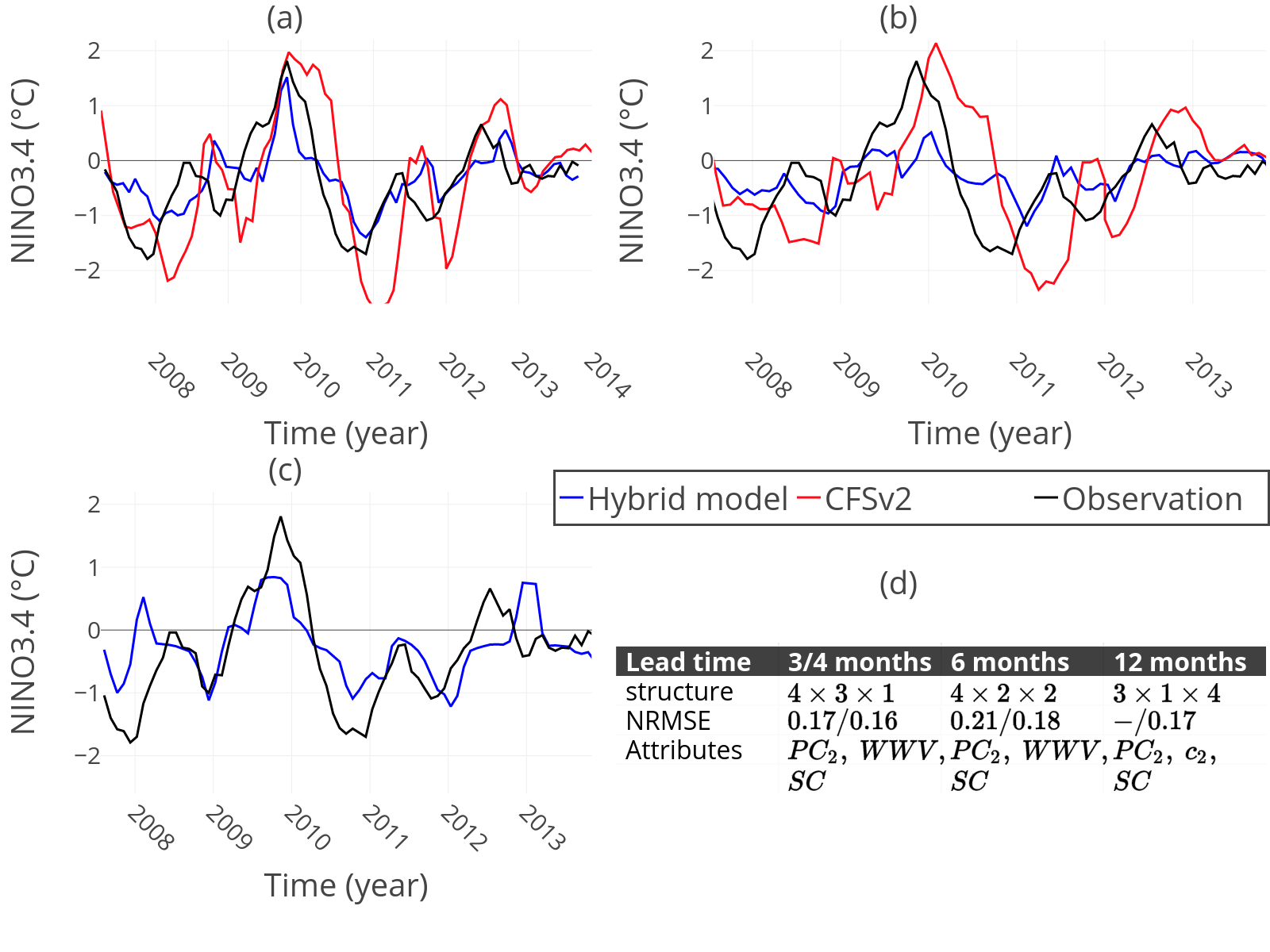}
      \caption{NINO3.4 predictions of the CFSv2 ensemble mean (red)
      and the hybrid model with ARIMA(12,1,0) (blue), compared to the observed index (black).
      For the hybrid model predictions, from an ensemble of 84 different
      ANN structures, structures resulting in a low NRMSE are presented. \textbf{(a)}~The 3-month
      lead time prediction of CFSv2 and 4-month lead time prediction of the hybrid model,
      \textbf{(b)}~the 6-month lead time predictions and \textbf{(c)}~12-month lead prediction.
      The CFSv2 ensemble does not predict 12~months ahead. \textbf{(d)}~Table containing information
      about all predictions: ANN structures of the hybrid model, NRMSEs of the CFSv2 ensemble
      mean and the hybrid model, and attributes used in the hybrid model predictions. \label{fig:lp}}
\end{figure*}

Considering the 3-, 6- and 12-month lead time predictions, both the
3- and 6-month lead time prediction of the CFSv2 ensemble show some lag
and amplification of the real NINO3.4 index (Fig.~\ref{fig:lp}). The hybrid
model predictions with ARIMA(12,1,0) resulting in a low NRMSE and relatively
simple ANN structure within an ensemble consisting of 84 different
ANN structures are also shown in Fig.~\ref{fig:lp}. The 84 different
structures are all structures up to three hidden layers with up to four
neurons.

Comparing the 3-month lead prediction of the CFSv2 ensemble with the
4-month lead prediction of the hybrid model, both the amplification and
the lag of the hybrid model prediction are smaller. While the lead time of
the hybrid model is 1~month longer, the prediction skill is better in terms
of NRMSE. The prediction skill of the hybrid model decreases at a 6-month
lead compared to the 4-month lead time prediction. Thereby the lag and
amplification of the CFSv2 prediction increase. Although the hybrid model
does not suffer as much from the lag, it underestimates the El~Niño event of
2010. In terms of NRMSE the hybrid model still obtains a better prediction
skill.

Although the shorter lead time predictions show slightly better results than
the conventional models, most important is a good prediction skill for larger
lead times that appears to overcome the spring predictability barrier. To
perform a 12-month lead prediction which could overcome this barrier, the
attributes from the shorter lead time predictions are found to be
insufficient. However, $c_2$ of the SSH network has shown to be predictive at
this lead time, according to its Granger causality and cross-correlation.
Therefore the WWV is replaced by $c_2$ for this prediction, which is related
to the same physical mechanism. In terms of NRMSE, the 12-month lead
prediction even improves the 6-month lead prediction of the hybrid model.
On average the prediction does not contain a lag in this period.

\begin{figure}[p]
\includegraphics[width=8.3cm]{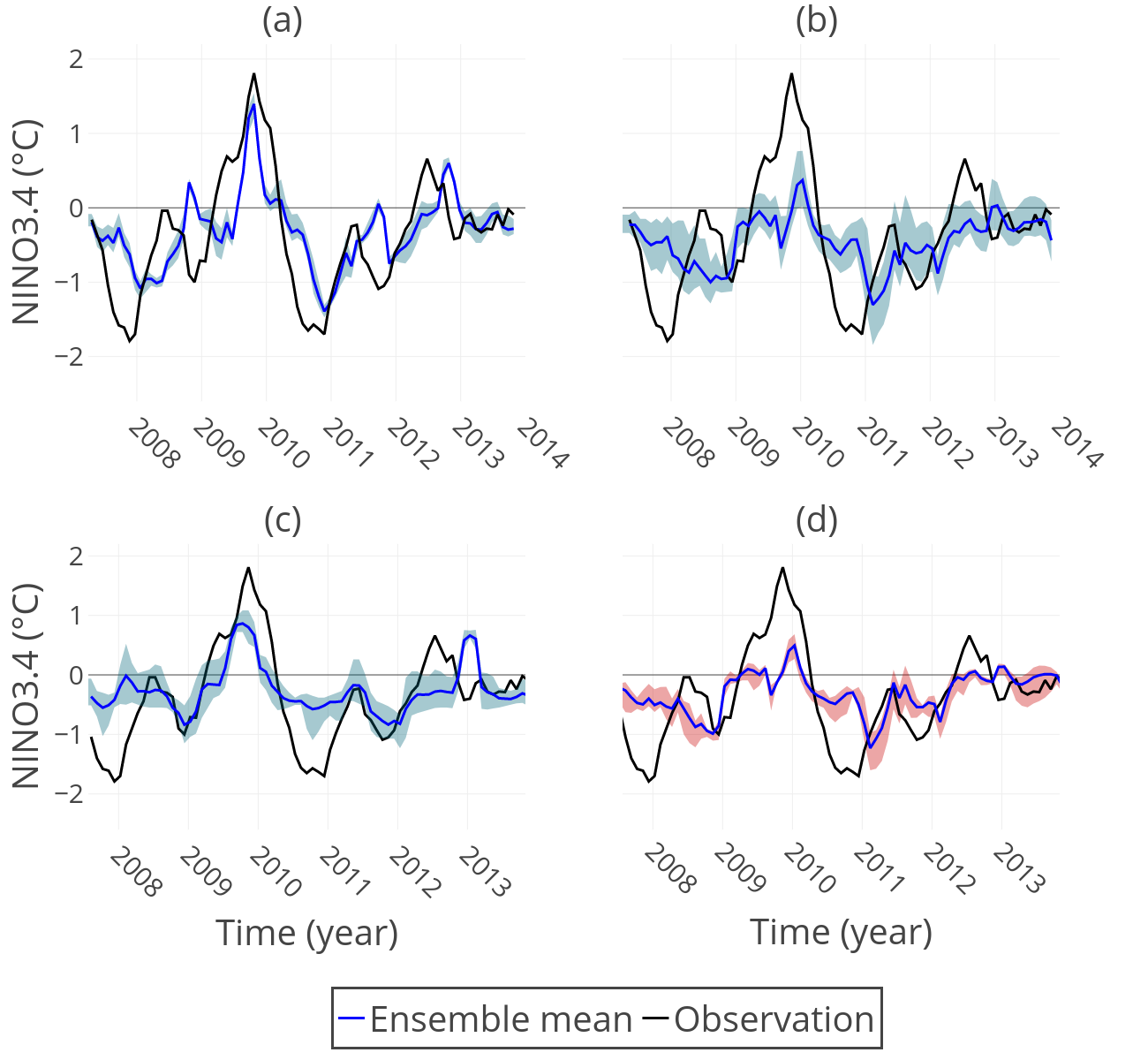}
\caption{Spread and mean (blue line) of ensembles of hybrid model predictions
with different hyperparameter values. The nine optimal (in terms of NRMSE)
predictions from the 84 different ANN structures at the
\textbf{(a)} 4-month lead time, \textbf{(b)} 6-month lead time and
\textbf{(c)} 12-month lead time. \textbf{(d)}~Ensemble with $9\leq p\leq
14$ in the ARIMA order with their optimal ANN structure at 6-month lead
time prediction (at the 4- and 12-month lead there is almost no
spread). Black is the observed NINO3.4 index.\label{fig:lpens}}
\end{figure}

\begin{figure}[p]
\includegraphics[width=8.5cm]{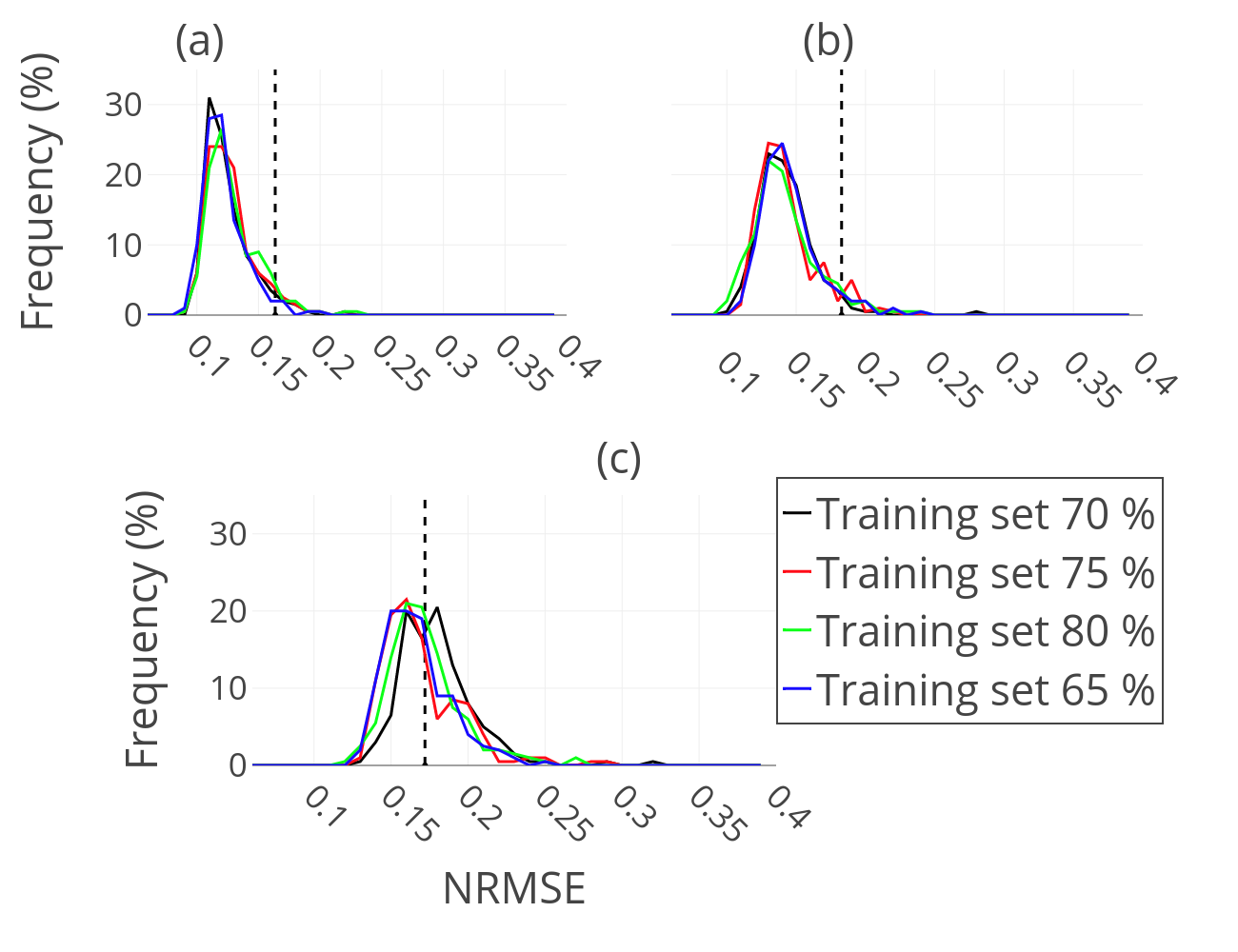}
\caption{Cross-validation results of the \textbf{(a)} 4-, \textbf{(b)} 6-
and \textbf{(c)}~12-month lead predictions of hybrid models from
Fig.~\ref{fig:lp}. Each line presents the frequency every NRMSE is obtained
for 200 different initial test sets with a specific training set/test set
percentage split. The vertical dashed line denotes the NRMSE of the
predictions of Fig.~\ref{fig:lp}.\label{fig:cv}}
\end{figure}

The hyperparameter values (i.e. the ARIMA order and the ANN structure) of the
predictions in Fig.~\ref{fig:lp} could still be a lucky shot. Therefore the
spread of the predictions with different hyperparameter values is shown in
Fig.~\ref{fig:lpens}. For the ANN structures, nine optimal (in terms of
NRMSE) predictions from the ensemble of 84 are considered. This
resulted in a higher spread in the 6- and 12-month lead prediction
compared to the 4-month lead prediction. For the ARIMA order all $9\leq
p\leq 14$ are chosen, which resulted in almost no spread for the 3- and
12-month lead prediction and a higher spread in the 6-month lead
prediction. Overall the models converge to similar predictions for those
different hyperparameter values.

\begin{figure}[t]
\includegraphics[width=8.3cm]{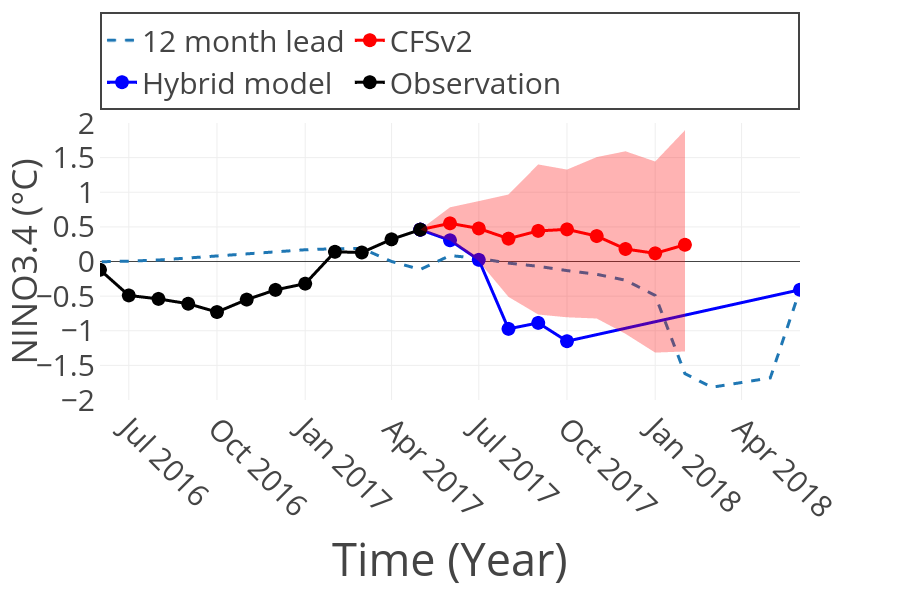}
\caption{NINO3.4 prediction from May 2017. In black the observed index until May 2017.
Red is the CFSv2 ensemble prediction mean and the shaded area is the spread of
the ensemble. The hybrid model prediction in blue is given by predictions from
hybrid models found to be most optimal at the different lead times with ARIMA(12,1,0).
The dashed blue line is the running 12-month lead time prediction.\label{fig:rpred}}
\end{figure}

To test the robustness of these results, a series of cross-validations has
been performed on the prediction models of Fig.~\ref{fig:lp}. Several
percentage splits have been chosen for the training and test set (65--35,
70--30, 75--25 and 80--20), but 200 different initial times of the test set
$t_i^\mathrm{test}$ are randomly chosen between March 1985 and December 2014.
This implies that $t_i^\mathrm{test}>t_f^\mathrm{train}$ is not necessarily
satisfied anymore. This allows us to make full use of the short time series
we have \citep{Bergmeir2012}. If the results for different training and test
sets do not deviate much, it is evidence that the model can generalize to an
arbitrary training and test set. The different percentage splits are chosen
since the size of a training set could possibly have an influence on the
prediction model. The cross-validation results of the hybrid models of
Fig.~\ref{fig:lp} are presented in Fig.~\ref{fig:cv}. At all three prediction
lead times, the peaks coincide at the same NRMSE for different training--test
set ratios. Therefore the different sizes of training and test sets do not
seem to influence the result. However, the width of the peaks increases when
the prediction lead time increases. This implies the prediction skill becomes
more sensitive to the choice of the training and test set with higher lead
time. Interestingly, at the 4- and 6-month lead time predictions, the
average NRMSE is lower than the NRMSE of the prediction of Fig.~\ref{fig:lp}.
This implies the predictions with a different training and test set are on
average even better than the prediction shown in Fig.~\ref{fig:lp}.

Finally, a prediction is made for the coming year in Fig.~\ref{fig:rpred}.
Different hybrid models are used at different lead times with ARIMA(12,1,0).
ANN structures are chosen that are found to be optimal at the different lead
times. For the predictions up to 5~months, the attributes WWV, PC$_2$ and
the SC are used from 1980 until the present. For the 12-month
lead prediction, the WWV is replaced by $c_2$ again. This time $c_2$ is
computed from the SSALTO/DUACS dataset. Therefore, only a dataset from 1993
until the present has been used to train the model and perform the 12-month
lead prediction.

Interestingly, as can be seen in Fig.~\ref{fig:rpred}, the hybrid model
typically predicts much lower ENSO development than the CFSv2 ensemble. The
uncertainty in the CFSv2 ensemble is large, since the spread of predictions
is between a strong El~Niño (NINO3.4 index between 1.5 and 2) and a moderate
La~Niña (NINO3.4 index between $-$1 and $-$1.5) for the coming 9~months. The
hybrid models predict development to a strong La~Niña (NINO3.4 index lower
than $-$1.5) the coming year. From the time of writing, only time will tell
which prediction is better. By the time of submission in early March 2018, La~Niña conditions are present according to the Climate Prediction Centre of
NCEP.

\conclusions[Summary and discussion\label{secconc}]

A successful attempt was made in this paper to use machine learning (ML)
techniques in a hybrid model to improve the skill of El~Ni\~no predictions.
Crucial for the success of this hybrid model is the choice of the attributes
applied to the artificial neural network. Here, we have explored the use of
network variables as additional attributes to several physical ones. Results
of the ZC model provided several interesting network variables. Of these
network variables, $c_2$ the amount of clusters of size two in a sea surface height (SSH) network
constructed from observations, is found to provide a warning of a
percolation-like transition in the SSH network. This percolation-like
transition coincides with an El~Niño event. This variable relates to the WWV
and hence the recharge/discharge mechanism, but extends the prediction lead
time of the WWV when applied in the prediction scheme. Furthermore, apart
from both these quantities related to ``recharge/discharge'', the PC$_2$ and the
seasonal cycle (SC) improve the prediction skill, representing respectively the
WWBs and the phase locking of ENSO. The flexibility of implementing different
variables at different lead times allows the hybrid model to improve on the
CFSv2 ensemble at short lead times (up to 6~months). Furthermore, it had a
better prediction result than all members of the CFSv2 ensemble in January
2010.

By including the network variable $c_2$, we obtained a 12-month lead time
prediction with comparable skill to the predictions at shorter lead times.
This prediction shows a step towards beating the spring predictability
barrier. Using ML has the advantage of recognizing the early warning signal
of $c_2$ as either a false or true positive. Therefore, it can be a more
reliable method then considering a warning when the signal exceeds a certain
threshold \citep{Ludescher2014VeryNino.}. Moreover, the early signal from the
network variable is not only used to predict an El~Niño event, but the
development of ENSO, as the hybrid model provides a regression of the NINO3.4
index. ML serves as a tool which is able to recognize important, but subtle,
patterns. Something the conventional statistical and dynamical models fail to
do in the chaotic system. In the end, the predictions from May 2017 are
discussed. By the time of writing, this is the prediction for the coming
year. The CFSv2 ensemble mean predicts neutral conditions for the coming 9~months,
with the spread between different members ranging from a strong El~Niño to a moderate
La~Niña. The hybrid model predicts moderate to strong La~Niña conditions for the coming year.

Although the results of the methods are promising, some adaptations to the
methods which select attributes could still improve predictions. Several
network variables resulted in a clear signal in the ZC model, but not
necessarily for the observations. Perhaps the cross-correlation and a Granger
causality test are not enough to determine the suitability of a variable in
the observations. Testing all possible attribute sets in the prediction
scheme and comparing results costs time. As a solution, the nonlinear methods
``lagged mutual information'' and ``transfer entropy'' can be techniques to
select variables at different lead times. After all, the attributes are
applied in the nonlinear part of the prediction scheme. Consequently, more
variables might be found to increase the prediction skill.

Even though the currently applied network measures showed interesting
properties, different climate network construction methods can still be
interesting to apply. The Pearson correlation is a simple, effective method
to define links between nodes. However, different properties of climate
networks could be found when using mutual information instead. Moreover, the
effect of spatial distance between nodes can be investigated and corrected
for \citet{Berezin2012StabilityTime}. Besides, we have limited ourselves to
networks within the Pacific area itself. As ENSO is an important mode in the
whole climate system, the area used for network construction might as well be
extended. More specifically, it can be interesting to include the Indian
Ocean in the network construction. Evidence is found that a cold SST to the
west of the Indian Ocean is related to a WWB a few months later
\citep{Wieners2016CoherentVariability}. This could result in a variable
related to WWBs, but increasing the lead compared to PC$_2$, which is
comparable to $c_2$ increasing the lead compared of the WWV.

By applying the ARIMA as a simple yet effective statistical method to apply in
the first step of the scheme, the hybrid model shows promising results.
However, the exact reason for how this model works remains a topic of
investigation. The ARIMA prediction could be related to the linear wave
dynamics. It can be interesting to replace the ARIMA part of the scheme by a
dynamical model accounting for these linear wave dynamics. For the same
reason, vector autoregression can be used instead of ARIMA. Being a
multivariate generalization of an autoregressive model, this can implement
the linear effect of other variables on ENSO.

Next to investigation of the exact reason the hybrid model works, some
adaptations could still improve the prediction scheme. For example, it is
assumed the linear and nonlinear part of the model are additive (see
Eq.~\ref{eq:add}). This is not necessarily the case for the real system
\citep{Khashei2011AForecasting}. Besides, the current model does not take
into account possible nonlinear effects from the history, since the ANN
describes a nonlinear function which does not depend on the history. The ANN
probably succeeds here because of its performance for nonlinear time series
in general. However, it could be interesting to investigate whether climate
network properties comprise enough of the nonlinear dynamics by themselves,
by combining them with a purely linear model. Moreover, the applied methods
searched for a prediction model which is most optimal in terms of least
squares minimization. However, it could be interesting to put larger weight
on predicting the extreme events in the optimization scheme (as the 6-month
lead predictions missed the 2010 El~Ni\~no event in Fig.~\ref{fig:lpens}), or
find a function which is simpler (e.g. applying a support vector machine
instead of ANN; \citealp{Pai2005AForecasting}).

A general difficulty in El~Ni\~no prediction is the short available
observational time series, also in other statistical prediction models
\citep{Drosdowsky2006}. Although different hyperparameters (the ANN structure
and ARIMA order) converge to a similar prediction and the prediction models
perform well at different training and test sets, the short time series makes
it difficult to perform another cross-validation method which completely
rules out that the model is overfitting.

\hack{\newpage}
Although the hybrid model and the attribute selection can clearly be
improved, the results here have shown the potential for ML methods, in
particular with network attributes, for El~Ni\~no prediction. The underlying
reason for this success is likely that through the network attributes, more
global correlations are taken into account which are needed to be able to
overcome the spring predictability barrier.

\hack{\clearpage}

\appendix
\section{}

This appendix summarizes the methods to calculate climate network properties.
The methods improved the prediction in the ZC model, but not for observational data.
Thus, they are not discussed in the main text. Appendix~\ref{sec:a1} defines
the different quantities and Appendix~\ref{sec:a2} their application to the ZC
model.

\subsection{Alternative network methods \label{sec:a1}}

From the unweighted network we
compute the local degree $d_i$ of node $i$ in the network as
\begin{equation}
d_i=\sum_jA_{ij} ,
\end{equation}
i.e. degree $d_i$ is equal to the amount of nodes that are connected to node $i$.

The spatial symmetry of the degree distribution is of interest, since it informs where
most links of the network are located. More specifically, our interest will be in the symmetry
in the zonal direction in a network. Therefore, the skewness of the meridional mean of
the degree in the network is calculated. This defines the zonal skewness of the degree
distribution in a network.

The following two climate network properties are derived from a so-called
NetOfNet approach. This is a network constructed with the same methods as
previously, but using multiple variables at each grid point (as specified in
Appendix~\ref{sec:a2}). This gives a network consisting of the networks from
the different variables interacting with each other. Only NetOfNet of two
different variables are considered. First, the cross clustering contains
information about the interaction between two unweighted networks. The local
cross clustering of a node is the probability that two connected nodes in the
other network are also connected to each other. The global cross clustering
$C_{vw}$ is the average over all nodes in subnetwork $G_v$ of the cross
clustering between $G_v$ and $G_w$:
\begin{equation}
C_{vw}=\frac{1}{N_v}\sum_{r}\frac{1}{k_r\left(k_r-1\right)}\sum_{p\neq q}A_{rp}A_{pq}A_{qr} .
\end{equation}
Here $r$ is a node in subnetwork $G_v$ of size $N_v$, $p$ and $q$ are the
nodes in the other subnetwork $G_w$, and $k_r$ denotes the cross degree of
node $r$ (i.e. amount of cross links node $r$ has with the other subnetwork).

The second NetOfNet property is the algebraic connectivity. This is the
second smallest eigenvalue ($\lambda_2$) of the Laplacian matrix as in
\cite{Newman2010Networks:Introduction} and describes the ``diffusion'' of
information in the network. In general, $\lambda_2>0$ if the network has a
single component.

A final network property $\Delta$ makes use of a differently calculated network
which is also undirected, but weighted. To construct it, the
cross-correlation $C_{ij}(\Delta t)$ at lag $\Delta t$, i.e. the Pearson
correlation between the variables $p_i(t)$ and $p_j(t+\Delta t)$ is
considered. Then the weights between the nodes are calculated by
\begin{equation}
W_{ij}=\frac{\max_{\Delta t}(C_{ij})-\text{mean}(C_{ij})}{\text{SD}(C_{ij})}\label{eq:weights} .
\end{equation}
Here $\max_{\Delta t}$ denotes the maximum, $\text{SD}$ the standard
deviation and $\text{mean}$ the mean value over all time steps that are
considered.

To calculate the property $\Delta$ of the network, links are added to a
network one by one, adding the link with the largest weight first
(Eq.~\ref{eq:weights}). At every step $T$ that a link is added, the size of
the largest cluster $S_1(T)$ is calculated. At the point of the percolation
transition, $S_1(T)$ increases rapidly. The size of this jump is $\Delta$:
\begin{equation}
\Delta=\max\left[S_1(2)-S_1(1),_{\cdots},S_1(T+1)-S_1(T),_{\cdots}\right] .\label{eq:Delta}
\end{equation}
The quantity $\Delta$ can be used to capture the percolation-like transition
\citep{Meng2017PercolationConditions}.

\begin{figure}[t]
      \includegraphics[width=8.3cm]{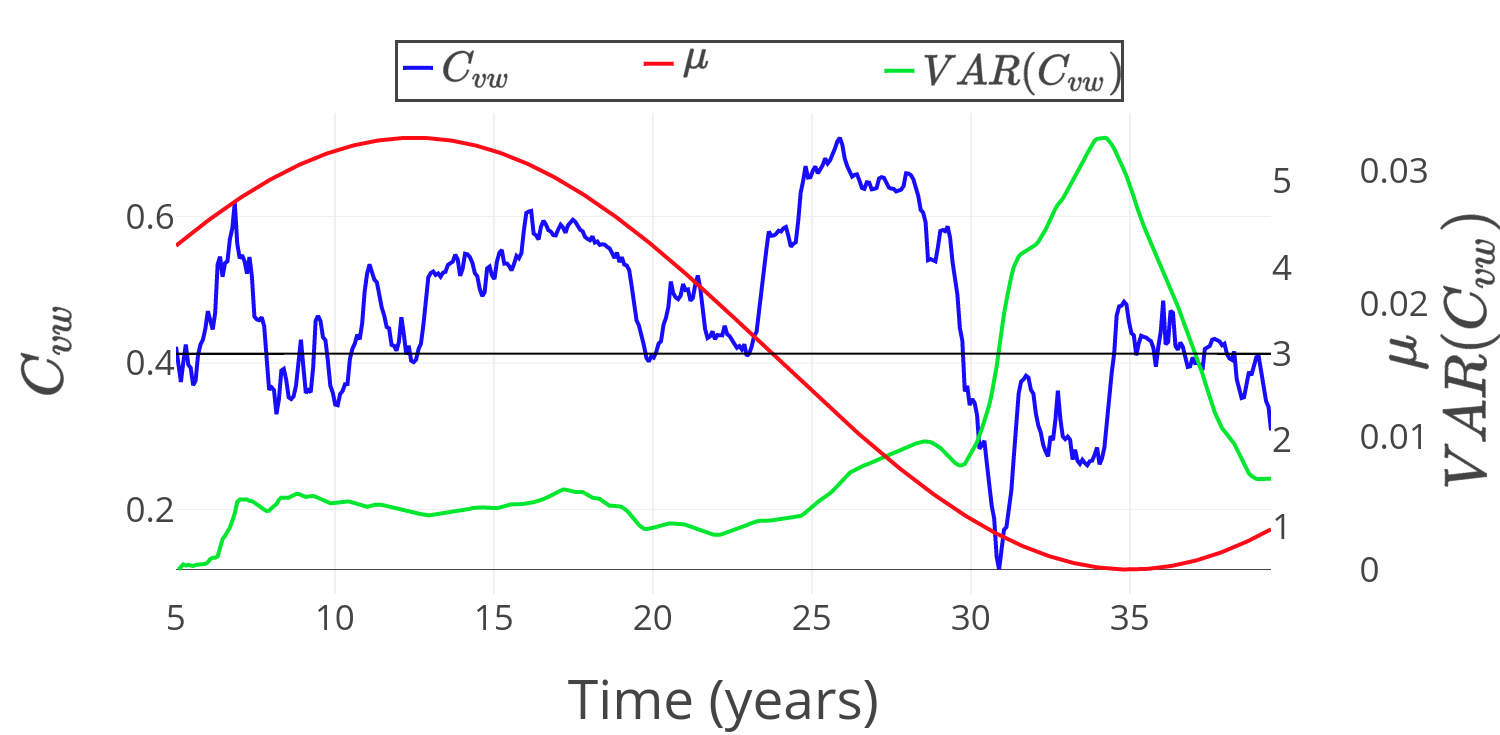}
      \caption{Global cross clustering between the SST and wind-stress network
      in blue and its variance in green in the ZC model. The coupling strength
      $\mu$ defined as a sinusoid around $\mu_\mathrm{c}=3$ with an amplitude
      of $0.25$ is in red. The sliding window is applied with a window of 5~years. }
      \label{fig:cross_clus}
\end{figure}

\begin{figure*}[t]
      \includegraphics[width=11cm]{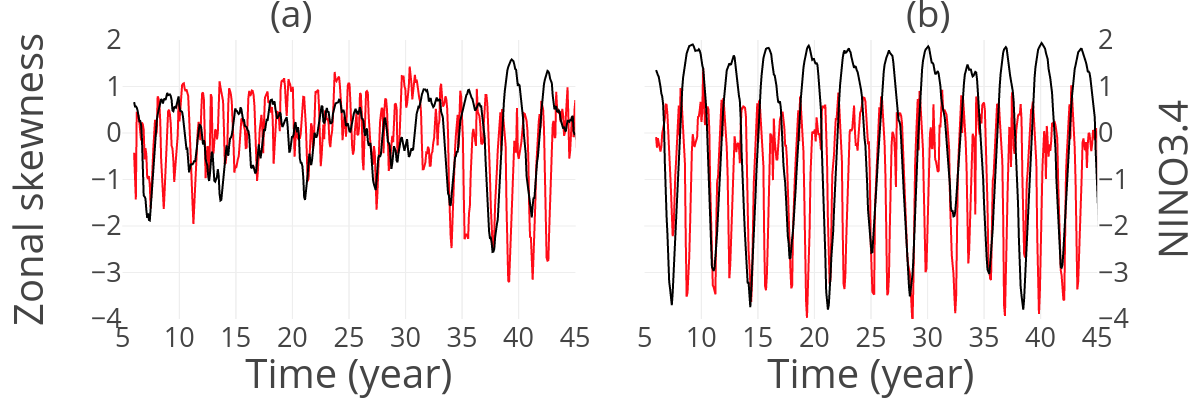}
      \caption{Zonal skewness of the degree field of the thermocline network with
      $\epsilon=0.6$ and a sliding window of 1~year in red, NINO3.4
      index in black in the ZC model. \textbf{(a)}~The subcritical ($\mu=2.7$)
      and \textbf{(b)}~the supercritical ($\mu=3.25$) case.}
      \label{fig:zS}
\end{figure*}

\subsection{Climate network properties of the ZC model \label{sec:a2}}

Determining how strong noise can excite the ENSO mechanisms in the
subcritical case, or determining whether the feedbacks sustain an
oscillation in the supercritical state, could provide information to increase
the prediction skill. \cite{Feng2015AVariability} found that the skewness of
the degree distribution $S_\mathrm{d}$ of the network reconstructed from SST
decreases monotonically with increasing coupling strength $\mu$. Although
$S_\mathrm{d}$ relates to the climate stability and coupling strength, it
does not inform whether the system is in either the supercritical or
subcritical state.

Here, we introduce a NetOfNet variable which may represent properties of the
stability of the background state: the global cross clustering ($C_{vw}$)
between the SST and wind-stress network. A sliding window of 5~years with
$\epsilon=0.6$ was used to compute the networks. In this case, the global
cross clustering coefficient is a measure of the amount of triangles in the
networks, containing one wind node and two SST nodes. In
Fig.~\ref{fig:cross_clus}, this cross clustering is calculated from data from
the ZC model, when coupling strength $\mu$ changes periodically in time
around the critical value $\mu_\mathrm{c} \sim 3.0$. Under subcritical
conditions, the noise has a larger influence on local correlations. This
causes triangles to break and the variance in the cross clustering
coefficient to increase. The cross clustering $C_{vw}$ is hence a diagnostic
network variable which informs whether the state of the system is in the
supercritical or subcritical regime.

Second, from the classical view of the oscillatory behaviour of ENSO, waves in
the thermocline should contain memory of the system, because of their
negative delayed feedback. The changing structure of the thermocline network
is therefore of interest when predicting ENSO. Calculating this network with
threshold $\epsilon=0.6$ and a sliding window with a length of 1~year, a
zonal pattern in the change of the network close to the equator can be
observed during an ENSO cycle. To compare network structures in the super-
and subcritical state, now constant $\mu=2.7$ (subcritical) and $\mu=3.25$
(supercritical) are taken. Generally, the degree field is quite spatially
symmetric, but when the ENSO turns either from upward to downward, or from
downward to upward, the degree of the nodes in the east decreases. This is at
the peak El~Niño or La~Niña.

\hack{\newpage}
To capture this zonal asymmetry around the equator with a variable, the zonal
skewness of the degree field will be used between $7{\degree}$\,S to
$7{\degree}$\,N. The higher the skewness, the more the degree will be located
west of the basin. If the skewness is close to zero, the degree is
symmetrically distributed over the basin. If it is low, most of the degree is
situated in the east. The skewness will show a negative peak when the ENSO
index is at its highest or lowest point in the cycle (Fig.~\ref{fig:zS}). In
the supercritical case $\mu=3.25$ this effect is indeed observed.
Nevertheless, in the subcritical case, the pattern is only visible once the ENSO
index shows a clear oscillation (around year 32).

Third, the quantity $\Delta$ behaves similar to $c_2$, when calculated from
the same (thermocline) network. Although $\Delta$ does not depend on a chosen
threshold like $c_2$, it peaks closer to an El~Niño event.

Finally, the algebraic connectivity ($\lambda_2$) can show the spread of
information within a network. Specifically, when considering an unweighted
NetOfNet from thermocline depth ($h$) and zonal wind ($\tau^x$) with
threshold $\epsilon=0.6$. The spread of information is relatively high before
an event, but also after an event, such that $\lambda_2$ peaks both before
and after an El~Niño event (both for $\mu=2.7$ and $\mu=3.25$).

\hack{\clearpage}

\dataavailability{All used observational data are from third parties and are either cited or can be found by URL as specified in Sect.~2.1. }

\competinginterests{The authors declare that they have no conflict of
interest.}

\begin{acknowledgements}
Peter~D.~Nooteboom would like to thank the Instituto de Física
Interdisciplinar y Sistemas Complejos (IFISC), for hosting his stay in
Mallorca during part of 2017.

Crist\'{o}bal~L\'{o}pez and Emilio~Hern\'{a}ndez-Garc\'{\i}a acknowledge
support from Ministerio de Economia y Competitividad and Fondo Europeo de
Desarrollo Regional through the LAOP project (CTM2015-66407-P,
MINECO/FEDER)\hack{\newline} \hack{\newline} Edited by: Ben
Kravitz\hack{\newline} Reviewed by: Robert Link and one anonymous referee
 \end{acknowledgements}


\begin{thebibliography}{58}
\providecommand{\natexlab}[1]{#1}
\providecommand{\url}[1]{{\tt #1}}
\providecommand{\urlprefix}{URL }
\expandafter\ifx\csname urlstyle\endcsname\relax
  \providecommand{\doi}[1]{doi:\discretionary{}{}{}#1}\else
  \providecommand{\doi}{doi:\discretionary{}{}{}\begingroup
  \urlstyle{rm}\Url}\fi

\bibitem[{Akaike(1974)}]{Akaike1974AIdentification}
Akaike, H.: {A New Look at the Statistical Model Identification}, IEEE
  Transactions on Automatic Control, AC-19, 716--723,
  \doi{doi:10.1109/TAC.1974.1100705}, 1974.

\bibitem[{Al-Smadi and Al-Zaben(2005)}]{Al-Smadi2005ARMAPerspective}
Al-Smadi, A. and Al-Zaben, A.: {ARMA Model Order Determination Using Edge
  Detection: A New Perspective}, Circuits, Systems Signal Processing, 24,
  723--732, 2005.

\bibitem[{Aladag et~al.(2009)Aladag, Egrioglu, and
  Kadilar}]{Aladag2009ForecastingMethodology}
Aladag, C.~H., Egrioglu, E., and Kadilar, C.: {Forecasting nonlinear time
  series with a hybrid methodology}, Applied Mathematics Letters, 22,
  1467--1470, \doi{10.1016/j.aml.2009.02.006}, 2009.

\bibitem[{Berezin et~al.(2012)Berezin, Gozolchiani, Guez, and
  Havlin}]{Berezin2012StabilityTime}
Berezin, Y., Gozolchiani, A., Guez, O., and Havlin, S.: {Stability of Climate
  Networks with Time}, Scientific Reports, 2, 1--8, \doi{10.1038/srep00666},
  2012.

\bibitem[{Bishop(2006)}]{Bishop2006PatternLearning}
Bishop, C.~M.: {Pattern Recognition and Machine Learning}, Springer-Verlag New
  York, 2006.

\bibitem[{Bjerknes(1969)}]{Bjerknes1969AtmosphericPacific}
Bjerknes, J.: {Atmospheric Teleconnections From The Equatorial Pacific},
  Monthly Weather Review, 97, 163--172,
  \doi{10.1175/1520-0493(1969)097<0163:ATFTEP>2.3.CO;2}, 1969.

\bibitem[{Bosc and Delcroix(2008)}]{Bosc2008ObservedOcean}
Bosc, C. and Delcroix, T.: {Observed equatorial Rossby waves and ENSO-related
  warm water volume changes in the equatorial Pacific Ocean}, Journal of
  Geophysical Research, 113, 1--14, \doi{10.1029/2007JC004613}, 2008.

\bibitem[{Bunge and Clarke(2014)}]{Bunge2014OnENSO}
Bunge, L. and Clarke, A.~J.: {On the Warm Water Volume and Its Changing
  Relationship with ENSO}, Journal of Physical Oceanography, 44, 1372--1385,
  \doi{10.1175/JPO-D-13-062.1}, 2014.

\bibitem[{Chen et~al.(2004)Chen, Cane, Kaplan, Zebiak, and
  Huang}]{Chen2004PredictabilityYears.}
Chen, D., Cane, M.~A., Kaplan, A., Zebiak, S.~E., and Huang, D.:
  {Predictability of El Ni{\~{n}}o over the past 148 years.}, Nature, 428,
  733--736, \doi{10.1038/nature02439}, 2004.

\bibitem[{Deza et~al.(2014)Deza, Masoller, and
  Barreiro}]{Deza2014DistinguishingNetworks}
Deza, J.~I., Masoller, C., and Barreiro, M.: {Distinguishing the effects of
  internal and forced atmospheric variability in climate networks}, Nonlinear
  Processes in Geophysics, 21, 617--631, \doi{10.5194/npg-21-617-2014}, 2014.

\bibitem[{Dijkstra(2006)}]{Dijkstra2006TheMechanisms}
Dijkstra, H.~A.: {The ENSO phenomenon: theory and mechanisms}, Advances in
  Geosciences, 6, 3--15, \doi{10.5194/adgeo-6-3-2006}, 2006.

\bibitem[{Fedorov et~al.(2003)Fedorov, Harper, Philander, Winter, and
  Wittenberg}]{Fedorov2003HowNino}
Fedorov, A.~V., Harper, S.~L., Philander, S.~G., Winter, B., and Wittenberg,
  A.: {How predictable is El Ni{\~{n}}o?}, Bulletin of the American
  Meteorological Society, 84, 911--919, \doi{10.1175/BAMS-84-7-911}, 2003.

\bibitem[{Feng(2015)}]{Feng2015AVariability}
Feng, Q.~Y.: {A complex network approach to understand climate variability},
  Ph.D. thesis, Utrecht University, 2015.

\bibitem[{Feng and Dijkstra(2016)}]{Feng2016ClimateVariability}
Feng, Q.~Y. and Dijkstra, H.~A.: {Climate Network Stability Measures of El
  Ni{\~{n}}o Variability}, 035801, \doi{10.1063/1.4971784}, 2016.

\bibitem[{Feng et~al.(2016)Feng, Vasile, Segond, Gozolchiani, Wang, Abel,
  Havlin, Bunde, and Dijkstra}]{Feng2016ClimateLearn:Measures}
Feng, Q.~Y., Vasile, R., Segond, M., Gozolchiani, A., Wang, Y., Abel, M.,
  Havlin, S., Bunde, A., and Dijkstra, H.~A.: {ClimateLearn: A machine-learning
  approach for climate prediction using network measures}, Geoscientific Model
  Development Discussions, pp. 1--18, \doi{10.5194/gmd-2015-273}, 2016.

\bibitem[{Fountalis et~al.(2015)Fountalis, Bracco, and
  Dovrolis}]{Fountalis2015ENSOCentury}
Fountalis, I., Bracco, A., and Dovrolis, C.: {ENSO in CMIP5 simulations:
  network connectivity from the recent past to the twenty-third century},
  Climate Dynamics, 45, 511--538, \doi{10.1007/s00382-014-2412-1},
  \urlprefix\url{http://dx.doi.org/10.1007/s00382-014-2412-1}, 2015.

\bibitem[{Gill(1980)}]{Gill1980SomeCirculation}
Gill, A.: {Some simple solutions for heat-induced tropical circulation}, Quart.
  J. Roy Meteor. Soc.,, 106, 447--462, 1980.

\bibitem[{Goddard et~al.(2001)Goddard, Mason, Zebiak, Ropelewski, Basher, and
  Cane}]{Goddard2001CurrentPredictions}
Goddard, L., Mason, S., Zebiak, S., Ropelewski, C., Basher, R., and Cane, M.:
  {Current Approaches to seasonal-to-interannual climate predictions},
  International Journal of Climatology, 21, 1111--1152,
  \doi{10.1080/002017401300076036}, 2001.

\bibitem[{Gozolchiani et~al.(2008)Gozolchiani, Yamasaki, Gazit, and
  Havlin}]{Gozolchiani2008PatternEvents}
Gozolchiani, A., Yamasaki, K., Gazit, O., and Havlin, S.: {Pattern of climate
  network blinking links follows El Ni{\~{n}}o events}, EPL (Europhysics
  Letters), 83, 28\,005, \doi{10.1209/0295-5075/83/28005}, 2008.

\bibitem[{Gozolchiani et~al.(2011)Gozolchiani, Havlin, and
  Yamasaki}]{Gozolchiani2011EmergenceNetwork}
Gozolchiani, A., Havlin, S., and Yamasaki, K.: {Emergence of El Ni{\~{n}}o as
  an autonomous component in the climate network}, Physical Review Letters,
  107, 1--5, \doi{10.1103/PhysRevLett.107.148501}, 2011.

\bibitem[{Guyon and Elisseeff(2003)}]{Guyon2003AnSelection}
Guyon, I. and Elisseeff, A.: {An Introduction to Variable and Feature
  Selection}, Journal of Machine Learning Research, 3, 1157--1182,
  \doi{10.1016/j.aca.2011.07.027}, 2003.

\bibitem[{Hall(1999)}]{Hall1999Correlation-basedLearning}
Hall, M.~A.: {Correlation-based Feature Selection for Machine Learning}, Ph.D.
  thesis, The university of Waikato, 1999.

\bibitem[{Hibon and Evgeniou(2005)}]{Hibon2005ToCombinations}
Hibon, M. and Evgeniou, T.: {To combine or not to combine: Selecting among
  forecasts and their combinations}, International Journal of Forecasting, 21,
  15--24, \doi{10.1016/j.ijforecast.2004.05.002}, 2005.

\bibitem[{Huang et~al.(2015)Huang, Banzon, Freeman, Lawrimore, Liu, Peterson,
  Smith, Thorne, Woodruff, and Zhang}]{Huang2015ExtendedIntercomparisons}
Huang, B., Banzon, V.~F., Freeman, E., Lawrimore, J., Liu, W., Peterson, T.~C.,
  Smith, T.~M., Thorne, P.~W., Woodruff, S.~D., and Zhang, H.~M.: {Extended
  reconstructed sea surface temperature version 4 (ERSST.v4). Part I: Upgrades
  and intercomparisons}, Journal of Climate, 28, 911--930,
  \doi{10.1175/JCLI-D-14-00006.1}, 2015.

\bibitem[{Hush(2017)}]{Hush2017MachinePhysics}
Hush, M.~R.: {Machine learning for quantum physics}, Science, 355, 580, 2017.

\bibitem[{Jin(1997)}]{Jin1997AnModel}
Jin, F.-F.: {An Equatorial Ocean Recharge Paradigm for ENSO. Part II: A
  Stripped-Down Coupled Model}, Journal of the Atmospheric Sciences, 54,
  830--847, \doi{10.1175/1520-0469(1997)054<0830:AEORPF>2.0.CO;2}, 1997.

\bibitem[{Jin et~al.(1994)Jin, Neelin, and Ghil}]{Jin1994ElChaos}
Jin, F.-F., Neelin, D.~J., and Ghil, M.: {El Ni{\~{n}}o on the Devil's
  staircase: Annual Subharmonic Steps to Chaos}, Science, 264, 70--72,
  \doi{10.1126/science.264.5155.70}, 1994.

\bibitem[{Khashei and Bijari(2011)}]{Khashei2011AForecasting}
Khashei, M. and Bijari, M.: {A novel hybridization of artificial neural
  networks and ARIMA models for time series forecasting}, Applied Soft
  Computing Journal, 11, 2664--2675, \doi{10.1016/j.asoc.2010.10.015}, 2011.

\bibitem[{Latif et~al.(1988)Latif, Biercamp, and von
  Storch}]{Latif1988TheBursts}
Latif, M., Biercamp, J., and von Storch, H.: {The response of a Coupled
  Ocean-Atmosphere General Circulation Model to Wind Bursts}, Journal of the
  Atmospheric Sciences, 45, 1988.

\bibitem[{Legler and Brien(1988)}]{Legler19882.TOGA}
Legler, D.~M. and Brien, J. J.~O.: {2. Tropical Pacific Wind Stress Analysis
  for TOGA}, Intergovernmental Oceanographic Commission 11, 1988.

\bibitem[{Ludescher et~al.(2014)Ludescher, Gozolchiani, Bogachev, Bunde,
  Havlin, and Schellnhuber}]{Ludescher2014VeryNino.}
Ludescher, J., Gozolchiani, A., Bogachev, M.~I., Bunde, A., Havlin, S., and
  Schellnhuber, H.~J.: {Very early warning of next El Ni{\~{n}}o.}, Proceedings
  of the National Academy of Sciences of the United States of America, 111,
  2064--6, \doi{10.1073/pnas.1323058111}, 2014.

\bibitem[{Madden and Julian(1994)}]{Madden1994ObservationsReview}
Madden, R.~A. and Julian, P.~R.: {Observations of the 40–50-Day Tropical
  Oscillation—A Review}, Monthly Weather Review, 122, 814--837,
  \doi{10.1175/1520-0493(1994)122<0814:OOTDTO>2.0.CO;2}, 1994.

\bibitem[{Meng et~al.(2016)Meng, Fan, Ashkenazy, and
  Havlin}]{Meng2016PercolationConditions}
Meng, J., Fan, J., Ashkenazy, Y., and Havlin, S.: {Percolation framework to
  describe El Ni{\~{n}}o conditions}, Chaos, 27, 1--15,
  \doi{10.1063/1.4975766}, 2016.

\bibitem[{Moore and Kleeman(1999)}]{Moore1999StochasticOscillation}
Moore, A.~M. and Kleeman, R.: {Stochastic forcing of ENSO by the intraseasonal
  oscillation}, Journal of Climate, 12, 1199--1220,
  \doi{10.1175/1520-0442(1999)012<1199:SFOEBT>2.0.CO;2}, 1999.

\bibitem[{Newman(2010)}]{Newman2010Networks:Introduction}
Newman, M.: {Networks: An introduction}, vol.~6, Oxford university press,
  Oxford, \doi{10.1017/S1062798700004543}, 2010.

\bibitem[{Pai and Lin(2005)}]{Pai2005AForecasting}
Pai, P.-F. and Lin, C.-S.: {A hybrid ARIMA and support vector machines model in
  stock price forecasting}, Omega, 33, 497--505,
  \doi{10.1016/j.omega.2004.07.024}, 2005.

\bibitem[{Philander(1990)}]{Philander1990ElOscillation}
Philander, S.~G.: {El Nino, La Nina, and the Southern Oscillation}, vol.~46,
  International Geophysics Series, San Diego, 1990.

\bibitem[{Rayner et~al.(2003)Rayner, Parker, Horton, Folland, Alexander,
  Rowell, Kent, and Kaplan}]{Rayner2003GlobalCentury}
Rayner, N.~A., Parker, D.~E., Horton, E.~B., Folland, C.~K., Alexander, L.~V.,
  Rowell, D.~P., Kent, E.~C., and Kaplan, A.: {Global analyses of sea surface
  temperature , sea ice , and night marine air temperature since the late
  nineteenth century}, Journal of Geophysical Research, 108,
  \doi{10.1029/2002JD002670}, 2003.

\bibitem[{Rebert et~al.(1985)Rebert, Donguy, Eldin, and
  Wyrtki}]{Rebert1985RelationsOcean}
Rebert, J.~P., Donguy, J.~R., Eldin, G., and Wyrtki, K.: {Relations between sea
  level, thermocline depth, heat content, and dynamic height in the tropical
  Pacific Ocean}, Journal of Geophysical Research, 90, 11\,719,
  \doi{10.1029/JC090iC06p11719}, 1985.

\bibitem[{Rissanen(1978)}]{Rissanen1978ModellingDescription}
Rissanen, J.: {Modelling by the shortest data description}, Automatica, 14,
  465--471, 1978.

\bibitem[{Rodr{\'{i}}guez-M{\'{e}}ndez
  et~al.(2016)Rodr{\'{i}}guez-M{\'{e}}ndez, Egu{\'{i}}luz~M,
  Hern{\'{a}}ndez-Garc{\'{i}}a, and
  Ramasco}]{Rodriguez-Mendez2016Percolation-basedSystems}
Rodr{\'{i}}guez-M{\'{e}}ndez, V., Egu{\'{i}}luz~M, V.~M.,
  Hern{\'{a}}ndez-Garc{\'{i}}a, E., and Ramasco, J.~J.: {Percolation-based
  precursors of transitions in extended systems}, Scientific Reports, 6,
  29\,552, \doi{10.1038/srep29552}, 2016.

\bibitem[{Runge(2014)}]{Runge2014DetectingSystems}
Runge, J.~G.: {Detecting and Quantifying Causal Interactions from Time Series
  of Complex Systems}, Ph.D. thesis, Humboldt-Universit{\"{a}}t zu Berlin,
  2014.

\bibitem[{Silver et~al.(2016)Silver, Huang, Maddison, Guez, Sifre, van~den
  Driessche, Schrittwieser, Antonoglou, Panneershelvam, Lanctot, Dieleman,
  Grewe, Nham, Kalchbrenner, Sutskever, Lillicrap, Leach, Kavukcuoglu, Graepel,
  and Hassabis}]{Silver2016MasteringSearch}
Silver, D., Huang, A., Maddison, C.~J., Guez, A., Sifre, L., van~den Driessche,
  G., Schrittwieser, J., Antonoglou, I., Panneershelvam, V., Lanctot, M.,
  Dieleman, S., Grewe, D., Nham, J., Kalchbrenner, N., Sutskever, I.,
  Lillicrap, T., Leach, M., Kavukcuoglu, K., Graepel, T., and Hassabis, D.:
  {Mastering the game of Go with deep neural networks and tree search}, Nature,
  529, 484--489, \doi{10.1038/nature16961}, 2016.

\bibitem[{Steinhaeuser et~al.(2012)Steinhaeuser, Ganguly, and
  Chawla}]{Steinhaeuser2012MultivariateNetworks}
Steinhaeuser, K., Ganguly, A.~R., and Chawla, N.~V.: {Multivariate and
  multiscale dependence in the global climate system revealed through complex
  networks}, Climate Dynamics, 39, 889--895, \doi{10.1007/s00382-011-1135-9},
  2012.

\bibitem[{Stolbova et~al.(2014)Stolbova, Martin, Bookhagen, Marwan, and
  Kurths}]{Stolbova2014TopologyLanka}
Stolbova, V., Martin, P., Bookhagen, B., Marwan, N., and Kurths, J.: {Topology
  and seasonal evolution of the network of extreme precipitation over the
  Indian subcontinent and Sri Lanka}, Nonlinear Processes in Geophysics, 21,
  901--917, \doi{10.5194/npg-21-901-2014}, 2014.

\bibitem[{Sun et~al.(2014)Sun, Li, Liu, Chow, Sun, and
  Wang}]{Sun2014UsingSeries}
Sun, Y., Li, J., Liu, J., Chow, C., Sun, B., and Wang, R.: {Using causal
  discovery for feature selection in multivariate numerical time series},
  Machine Learning, \doi{10.1007/s10994-014-5460-1}, 2014.

\bibitem[{Tsonis et~al.(2006)Tsonis, Swanson, and
  Roebber}]{Tsonis2006WhatClimate}
Tsonis, A.~A., Swanson, K.~L., and Roebber, P.~J.: {What do networks have to do
  with climate?}, Bulletin of the American Meteorological Society, 87,
  585--595, \doi{10.1175/BAMS-87-5-585}, 2006.

\bibitem[{Tupikina et~al.(2014)Tupikina, Rehfeld, Molkenthin, Stolbova, Marwan,
  and Kurths}]{Tupikina2014CharacterizingNetworks}
Tupikina, L., Rehfeld, K., Molkenthin, N., Stolbova, V., Marwan, N., and
  Kurths, J.: {Characterizing the evolution of climate networks}, Nonlinear
  Processes in Geophysics, 21, 705--711, \doi{10.5194/npg-21-705-2014}, 2014.

\bibitem[{Tziperman et~al.(1994)Tziperman, Stone, Cane, and
  Jarosh}]{Tziperman1994ElOscillator}
Tziperman, E., Stone, L., Cane, M.~A., and Jarosh, H.: {El Nino chaos:
  Overlapping of resonances between the seasonal cycle and the pacific
  ocean-atmosphere oscillator}, Science, 264, 72--74,
  \doi{10.1126/science.264.5155.72}, 1994.

\bibitem[{Valenzuela et~al.(2008)Valenzuela, Rojas, Rojas, Pomares, Herrera,
  Guillen, Marquez, and Pasadas}]{Valenzuela2008HybridizationPrediction}
Valenzuela, O., Rojas, I., Rojas, F., Pomares, H., Herrera, L.~J., Guillen, A.,
  Marquez, L., and Pasadas, M.: {Hybridization of intelligent techniques and
  ARIMA models for time series prediction}, Fuzzy Sets and Systems, 159,
  821--845, \doi{10.1016/j.fss.2007.11.003}, 2008.

\bibitem[{van~der Vaart et~al.(2000)van~der Vaart, Dijkstra, and
  Jin}]{vanderVaart2000TheModel}
van~der Vaart, P. C.~F., Dijkstra, H.~A., and Jin, F.~F.: {The Pacific Cold
  Tongue and the ENSO Mode: A Unified Theory within the Zebiak–Cane Model},
  Journal of the Atmospheric Sciences, 57, 967--988,
  \doi{10.1175/1520-0469(2000)057<0967:TPCTAT>2.0.CO;2}, 2000.

\bibitem[{Von Der~Heydt et~al.(2011)Von Der~Heydt, Nnafie, and
  Dijkstra}]{VonDerHeydt2011ColdPliocene}
Von Der~Heydt, A.~S., Nnafie, A., and Dijkstra, H.~A.: {Cold tongue/Warm pool
  and ENSO dynamics in the Pliocene}, Climate of the Past, 7, 903--915,
  \doi{10.5194/cp-7-903-2011}, 2011.

\bibitem[{Wang et~al.(2015)Wang, Gozolchiani, Ashkenazy, and
  Havlin}]{Wang2015OceanicNetworks}
Wang, Y., Gozolchiani, A., Ashkenazy, Y., and Havlin, S.: {Oceanic
  El-Ni{\~{n}}o wave dynamics and climate networks}, New Journal of Physics,
  18, 1--5, \doi{https://doi.org/10.1088/1367-2630/18/3/033021}, 2015.

\bibitem[{Wieners et~al.(2016)Wieners, de~Ruijter, Ridderinkhof, von~der Heydt,
  and Dijkstra}]{Wieners2016CoherentVariability}
Wieners, C.~E., de~Ruijter, W.~P., Ridderinkhof, W., von~der Heydt, A.~S., and
  Dijkstra, H.~A.: {Coherent tropical Indo-Pacific interannual climate
  variability}, Journal of Climate, 29, 4269--4291,
  \doi{10.1175/JCLI-D-15-0262.1}, 2016.

\bibitem[{Wu et~al.(2006)Wu, Hsieh, and Tang}]{Wu2006NeuralTemperatures}
Wu, A., Hsieh, W.~W., and Tang, B.: {Neural network forecasts of the tropical
  Pacific sea surface temperatures}, Neural Networks, 19, 145--154,
  \doi{10.1016/j.neunet.2006.01.004}, 2006.

\bibitem[{Yeh et~al.(2009)Yeh, Kug, Dewitte, Kwon, Kirtman, and
  Jin}]{Yeh2009ElClimate}
Yeh, S.-W., Kug, J.-S., Dewitte, B., Kwon, M.-H., Kirtman, B.~P., and Jin,
  F.-F.: {El Ni{\~{n}}o in a changing climate}, Nature, 461, 511--514,
  \doi{10.1038/nature08316}, 2009.

\bibitem[{Zebiak and Cane(1987)}]{Zebiak1987AOscillation}
Zebiak, S.~E. and Cane, M.~A.: {A model El Ni{\~{n}}o-Southern Oscillation},
  Monthly Weather Review, 115, 2262--2278,
  \doi{10.1175/1520-0493(1987)115<2262:AMENO>2.0.CO;2}, 1987.

\bibitem[{Zhang(2003)}]{Zhang2003TimeModel}
Zhang, G.: {Time series forecasting using a hybrid ARIMA and neural network
  model}, Neurocomputing, 50, 159--175, \doi{10.1016/S0925-2312(01)00702-0},
  2003.

\end{thebibliography}
\end{document}